# 3D atomistic imaging of polymer nanocomposites with Atom Probe Tomography: experimental methodology, preliminary results and future outlook

James O. Douglas*,[1], Reza Salehiyan[2,a], Aparna Saksena[3], Tim M. Schwarz[3], Baptiste Gault[1,3,b], Stella Pedrazzini[1], Emilio Martinez-Paneda[4], Łukasz Figiel[2]

*Corresponding author. Department of Materials, Royal School of Mines, Imperial College London, LondonSW7 2AZ, , UK. Tel: +44 (0)20 7589 5111 Fax: +44 (0)20 7589 5111. Email: j.douglas@imperial.ac.uk

[1]Department of Materials, Royal School of Mines, Imperial College London, London, SW7 2AZ, UK.

[2]International Institute for Nanocomposites Manufacturing (IINM), WMG, University of Warwick, Coventry, CV4 7AL, UK.

[3]Microstructure Physics and Alloy Design, Max Planck Institute for Sustainable Materials (formerly known as Max-Planck-Institut für Eisenforschung GmbH), Max-Planck-Str. 1, Düsseldorf 40237, Germany.

[4]Department of Engineering Science, University of Oxford, Oxford, OX1 3PJ, UK.

[a]Current address: School of Computing, Engineering and the Built Environment, Edinburgh Napier University, Edinburgh EH10 5DT, , UK.

[b]Current address: Avenue de l'Université, 76800, Saint Etienne du Rouvray, France.

## Abstract

The use of polymer nanocomposites as gas barrier materials has seen increasing interest, including applications involving hydrogen transport and storage. Better understanding of gas transport through those polymeric systems requires 3D nanoscale detection of distributions and the possible trapping of gas molecules within nanoparticles and polymer/nanoparticle interfaces While atom probe tomography (APT) offers promising means for such nanoscale characterisation, its use for polymers has been mainly limited to thin organic layers deposited onto substrates or pre-fabricated metal needle shaped specimens. This work provides the very first application of APT to bulk polymer nanocomposites. Particularly, site specific atom probe sample preparation by Focused Ion Beam (FIB) liftout has been shown for the first time in a model system of hexagonal boron nanoparticles within a PVDF polymer matrix, using a variety of FIB workflows including Xe FIB, Ga FIB, cryogenic Ga FIB and deuterium charging. Mass spectra from the bulk polymer and the nanoparticle were collected using pulsed laser atom probe using standard conditions and compared. Several challenges encountered during this research including damage of the polymeric matrix during sample preparation were extensively discussed in this paper. Once those challenges have been resolved (e.g. by developing site specific sample preparation protocols), the application of APT to polymer nanocomposites can open new options for nanoscale characterisation of those systems.

## Introduction

Polymers reinforced with nanoscale particles (so called polymer nanocomposites) offer promising means for property enhancement such as mechanical, dielectric, thermal or electrical (conductivity) (Mu et al., 2018; Ellingford et al. 2018; Smith et al., 2020). Particularly, 2D nanoparticles have been found to be particularly useful in controlling gas transport by creating tortuous paths that limit diffusion of gas molecules through polymers (Unalan et al., 2015; Liu et al., 2024). This specific feature can promote further applications of those 2D-reinforced nanocomposites as separation membranes (e.g. fuel cells), or coatings/liners for gas-related infrastructure (e.g. storage tanks, or pipelines) (Figiel et al., 2025). The gas transport behavior of these systems will be heavily dependent on the size and distribution of the nanoparticles within the polymer matrix, along with the physical and chemical interactions that these particles have with the matrix and the gases of interest during transport. Multi-length scale characterization of these systems is required to model, optimise and predict their behavior, with nanoscale analysis of the matrix/particle interfaces required to better understand their interactions with gases.

While characterization techniques such as atomic force microscopy (AFM), Scanning and Transmission Electron Microscopy (SEM and TEM) are excellent experimental tools to provide information about the properties and morphology of the nanocomposites, there are challenges in quantifying the nanoscale distribution of low atomic number elements present in polymer systems such as hydrogen, oxygen and nitrogen and therefore additional techniques are required (Herbig, 2018; McGilvery et al., 2020). Additionally, quantitative bulk characterization techniques such as Thermal Desorption Spectroscopy (TDS) (Jung et al., 2021) are applied to study rate of gas desorption from a sample, and determine permeation characteristics (e.g. solubility, diffusivity, and permeability). Atom Probe Tomography (APT) is a 3D spatially resolved time-of-flight mass spectrometry technique with up to tens of ppm compositional sensitivity and up to 0.2 nm spatial resolution (in the analysis direction) within nanoscale volumes of approximately 50 nm x 50 nm x 100 nm (Gault et al., 2021). It can detect low atomic number elements such as hydrogen/deuterium (Chen et al., 2023; Gault et al., 2024), boron (Meisenkothen et al., 2015), beryllium (Jayaram et al., 1993), carbon (Sha et al., 1992) (Thuvander et al., 2019), oxygen (Zanuttini et al., 2017) and nitrogen (Sha et al., 1992) (Gault et al., 2016) (albeit with some element specific challenges in quantification) and so has potential application in analyzing polymer nanocomposite systems. During atom probe analysis, atoms are ionized and field evaporated (either by high voltage pulses or laser pulses) from the surface of a nanoscale needle shaped specimen with an apex diameter of less than 100 nm. These ions are identified in terms of mass/charge ratio by a mass spectrometer combined with a 2D sensitive detector plate, and the evaporated volume can be reconstructed in 3D by back-projecting these ions.

APT has specific applications in characterizing compositional changes in complex 3D nanoscale features such as grain boundaries (Jenkins et al., 2020), nanoscale phases (Bagot et al., 2017) and nanoparticles (Yang et al., 2019) which can be challenging to chemically and compositionally quantify using spectroscopic techniques such as STEM energy-dispersive X-ray (EDS)/electron energy-loss spectroscopy (EELS) (Herbig et al., 2015) (de Gabory et al., 2015) or scanning probe techniques such as Scanning Tunnelling Microscopy (Gajjela et al., 2021). Therefore, APT can be extended to be useful in the analysis of polymer nanocomposite systems in characterizing the chemical composition of their interfaces which frequently determine their macroscopic properties. For example, depending on the level of nanoparticle functionalization, or lack of it, nano-defects or polymer regions with modified chain dynamics can arise and affect both their local and global behaviour (Quaresimin et al., 2016). As this remains an open fundamental challenge in nanocomposites research, the APT approach could offer robust means for investigating structure-property relations in those polymeric systems. APT was initially restricted to conductive metals due to the limitations of sample preparation and instrumentation (Kelly & Miller, 2007) but with the combined introduction of laser pulsed induced field evaporation (Bunton et al., 2007) and Focused Ion Beam (FB) lift-out (Thompson et al., 2007), there has been a significant increase in the number of material systems that can be analyzed including semiconductors (Giddings et al., 2018), geological materials (Reddy et al., 2020), biological materials (Grandfield et al., 2022) and soft matter (Adineh et al., 2018).

The APT analysis of soft matter material systems such as polymers has been mainly limited by difficulties in preparing suitable nanoscale needle shaped specimens without causing damage from electron and ion beams during typical Focused Ion Beam (FIB) sample preparation. Therefore to analyze polymers with APT, the most viable method was found to be the deposition of polymer films directly onto pre-fabricated support structures (Nakagawa & Ishida, 1973; Nishikawa & Kato, 1986; Prosa et al., 2010). In other cases, where the aim is to analyse features embedded in resin rather than the resin itself, FIB milling (Perea et.,2016) or laser milling (Sharm et al., 2024) is used, where electron and ion beam damage can be minimized through reducing the temperature of the sample through cryogenic stages (Bassm et al, 2012) (Hayles & De Winter, 2021) (Parmenter & Nizamudeen, 2021) (Chang et al., 2019). Fully cryogenic sample preparation for atom probe has been established in recent years (Schreiber et al., 2018) (Douglas et al. 2023) and provides a new pathway towards analyzing organic/liquid systems (Woods et al., 2023). While APT samples are significantly more challenging to prepare at cryogenic temperatures compared to room temperature, cryogenic conditions can be exploited to prevent out diffusion of gas species after exposing specimens to gas charging (Takahashi et al., 2010) - the cryogenics approach has been used extensively in the characterization of hydrogen/deuterium segregation behavior in metals (Chen et al., 2017), semiconductors (Tweddle et al., 2021) and geological materials (Daly et al., 2021).

Secondly, there are also significant challenges in the interpretation of the mass spectra associated with APT analysis of organic materials such as polymers. The high electrostatic field conditions of atom probe analysis will lead to the fragmentation of the polymer chains into a variety of complex molecular ions instead of elemental ions (Maruyama et al., 1987), including multiple peaks which may only differ by a single Dalton. The complexity of these mass spectra makes the unambiguous quantification and identification of each peak difficult due to a high number of mass spectra overlaps between them (London, 2019). In the case where both the carbon-based polymer matrix and the nanoparticles are compositionally similar, such as in the case of graphene sheets in an organic matrix, it would be difficult to reliably distinguish between them.

There are considerable requirements for technical development in determining whether atom probe can be applied to the nanoscale study of polymer nanocomposites and the interfaces between the polymer bulk and nanoparticle fillers. To this end, an academically and industrially relevant model system is required to develop a robust workflow and address the specific challenges of sample preparation and analysis. This would require that both polymer matrix and nanoparticles are compositionally enough different from each other to be readily distinguished via mass spectra comparisons and to either have a high enough number density to be found randomly during sample preparation or for them to be visible during FIB site specific sample preparation.

This paper is the very first attempt to address some of the aforementioned challenges in APT application to polymer nanocomposites by developing sample preparation FIB workflows including room temperature and cryogenic conditions, as well as deuterium charging, using a model polymer nanocomposite system. Consequently, this work paves the way for future advances in this field and application of APT to study nanoscale phenomena in polymer nanocomposites.

## Materials and Methods

**Nanocomposite Fabrication**

A model polymer nanocomposite system comprising of polyvinylidene fluoride (PVDF) polymer matrix and hexagonal boron nitride (hBN) nanoparticles was fabricated for subsequent atom probe analysis. The polymer matrix and nanoparticles contained specific identifying elements (fluorine for the polymer and boron for the nanoparticles), which would help to distinguish the characteristic mass spectra peaks from each. Boron nitride nanoparticles with a length scale of a few hundred nm were selected to be large enough to be visible during FIB sample preparation to remove ambiguity as to whether there were nanoparticles within the final analyzed volume.

Material processing and physical characterization was carried out in the International Institute for Nanocomposites Manufacturing (IINM) at the University of Warwick. As received PVDF (Kynar 740, Arkema) pellets were cryogenically milled prior to melt compounding to enhance its mixing with 2D hexagonal boron nitride (hBN) (Elinova, Thomas Swan) nanoparticles. Subsequent melt compounding was carried out within a twin-screw extruder (Prism Eurolab 16mm, Thermo Fisher Scientific) at 190 ˚C and 100 rpm, and the resulting material was transferred as nanocomposite pellets for injection moulding of disc-like (d =~25 mm, t =~1.5 mm) specimens at 230 ˚C – the hBN loading in the nanocomposite varied between 0.1 wt. % and 10 wt. %.

**SEM analysis**

Injection moulded nanocomposite samples were cryogenically fractured in liquid nitrogen to expose fracture surfaces. The fracture surfaces were then sputter-coated using an Au/Pd target for fracture surface analysis by SEM (Zeiss Sigma, InLens detector at 1 – 2 kV). A 5 wt.% hBN loading was selected for subsequent APT sample preparation due to a more regular distribution of particles to ensure that a typical APT liftout lamella of 30 μm x 3 μm x 5 μm contained a sufficient number of nanoparticles for analysis.

A section of a disc-like specimen of the 5 wt.% hBN loaded material was cut with scissors to approximately 3 mm × 7 mm × 1.5 mm, and then mounted on a commercial atom probe clip support (CAMECA) and coated with a thin layer of conductive silver paint apart from a small region on the surface which was left bare to allow access for site specific FIB sample preparation. Silver paint was used instead of a thin metal sputtered layer as ion beam milling can often remove local areas of sputtered films and charging can become an issue over time.

Room temperature cross sectional FIB milling for SEM was carried out with 30 kV $Ga^+$ ions in a Thermo Fisher Scientific Helios 5 CX. A 100 nm thick layer of platinum compound was deposited by electron beam (2 kV and 2.8 nA) followed by a further 1 μm thick layer of platinum compound deposited by ion beam (30 kV $Ga^+$ and 0.46 nA). High resolution SEM images were collected using 2 kV and currents between 13 pA and 0.1 nA using a through-lens detector (TLD) in immersion mode in both secondary electron and backscatter electron modes.

SEM EDS was carried out on both the planar surface of the 5 wt.% hBN loading PVDF sample and a 50 μm wide Ga FIB cross section of the same using a Zeiss Auriga Crossbeam using an Oxford Instruments EDS detector at 5 mm WD using 5 kV, a 30 μm aperture and a 300s acquisition time. Planar EDS was carried out at 0 degree stage tilt using an array of randomised spot acquisitions as no clear particles were observable on the surface. EDS of the FIB cross section was carried out at 52 degree tilt using spot acquisitions on visible high contrast particles.

**Site Specific FIB based APT Sample Preparation**

Room temperature trenching, liftout and milling of atom probe samples was carried out using a Thermo Fisher Scientific Hydra 5CX 30 kV $Xe^+$ plasma in accordance with standard APT approaches (Thompson et al., 2007). Xe plasma was initially used to reduce sample preparation damage that is often associated with APT sample preparation using Ga FIB systems (Eder et al., 2021). A 100 nm thick layer of platinum compound was deposited by electron beam (2 kV and 2.8 nA) followed by a further 1 μm thick layer of platinum compound deposited by ion beam (12 kV $Xe^+$ and 0.33 nA). Trenching of the lamella was carried out with 30 kV $Xe^+$ ions between 4 nA and 0.33 nA, deposition and mounting using 12 kV 0.1 nA and sharpening using

30 kV 0.33 nA to 30 pA followed by a 5 kV polishing stage except in the single sample which was successfully fabricated with a boron nitride particle at the apex. The particle/bulk polymer interface was at risk of being removed entirely if a polishing stage was carried out due to issues in milling rates between the matrix and particle during the sharpening process - hence in this case the low kV polishing stage was omitted, which is the likely reason for Xe ions being present in the surface regions of the mass spectrum. Electron beam-induced damage in the form of sample bending and deformation in sharpened specimens was observed even during low dose SEM imaging at 2 kV and 50–100 pA at high magnification - thus imaging currents and dwell times were minimized as much as possible for both electrons and ions.

APT specimens from bulk polymer and a polymer region containing a boron nitride particle at the apex were then analyzed with laser pulsing mode using a Local Electrode Atom Probe (LEAP) (CAMECA) 5000 XR. The conditions for each dataset can be found on the mass spectra for the sample. As these materials have not been previously investigated, analysis conditions were varied during initial testing and conservative approaches were used with high base temperatures of 80 K used along with laser energies up to 50 pJ. The data was processed using the Integrated Visualisation and Analysis Software (IVAS) as part of the Atom Probe Suite 6.3 package (CAMECA). A shank angle reconstruction was used to aid in visualization of the reconstructed data, but this should not be regarded as a definitive physical reconstruction due to non-uniform evaporation of the material during atom probe analysis. Shank angle reconstructions enforce a consistent shank angle to the reconstruction based on a user provided angle and initial tip diameter. This is based on the assumption that the tip evaporates uniformly during analysis, which is known to be incorrect when in non-homogenous samples such as multilayered materials (Marquis et al., 2011), or embedded nanoparticles (Larson et al., 2015). Iso-concentration surfaces were used to delineate a region of high concentration species but were not coherent or smooth enough for further use of a concentration profile. Hence, 1D concentration profiles were used to more reliably capture the change in composition across interfaces.

Detector event maps and voltage curves showing the distribution of detected events and voltage evolution during analysis are shown in the Supplementary Information (Figures S1-S8).

The bulk polymer samples could be sharpened with little difficulty into standard sized and shaped atom probe needle shaped specimens within an end diameter of less than 100 nm. The boron nitride nanoparticles were visible in secondary electron imaging as high contrast features whilst they were similar in contrast to the low atomic number polymer bulk when imaged in backscattered electron imaging. Due to the non-uniform milling behavior of the nanoparticle/polymer interface, it was not possible to thin the sample containing the nanoparticle region to less than 100 nm. However as polymers may not always require as high field to evaporate when compared to metals as has been found when analyzing thin films on relatively large, prefabricated metal substrates (Proudian et al., 2019), therefore a larger sample could be considered preferable and in this case was found to be acceptable.

**Deuterium charging of APT samples**

To understand the gas barrier function and interaction of gas molecules with the nanoparticles and/or polymer matrix, deuterium gas charging using the ReactHub System at MPI SusMat of the pre-sharpened APT specimens was performed (Stephenson et al., 2018). Deuterium was used to facilitate the disambiguation in the mass spectra between hydrogen signals, which are always present in mass spectra due to residual H in the measurement chamber (Sundell et al., 2013), and the hydrogen located within the specimen as discussed in recent review articles (Chen et al., 2023) (Gault et al., 2024).

For the sample preparation, a standard lift-out method at room temperature was performed using a Ga FIB and Plasma-FIB. The specimens were sharpened under cryogenic temperatures using the Ga-FIB Helios 5CX equipped with a Aqillos cryogenic stage (Thermo Fisher Scientific) to minimize the beam damage. Cryogenic in-situ surface coating of finished atom probe samples using high purity Cr was carried out on some of the cryogenically sharpened Ga FIB samples to determine if an additional surface conductive layer would improve sample yield and data quality (Schwarz et al., 2024).

After the sample preparation, the specimens were transferred through a Vacuum Cryogenic Transfer Module (VCTM) (Ferrovac) to the ReactHub (Stephenson et al., 2018) gas reaction cell at room temperature and ultra-high-vacuum UHV ($10E^{-11}$ mbar). The deuterium charging was performed at 250 mbar $D_2$ at room temperature for ~ 6 hours as described in (Saksena et al., 2024). Afterwards the sample was cooled down to cryogenic temperatures by loading it into a cryogenically cooled VCTM stage and transferred to LEAP 5000 XS atom probe (CAMECA) for further analysis.

## Results

The hBN nanoparticles were observed to have a nominal 0.5 µm to 1 µm lateral size and a few layer/multi-layer structure, as shown by the exposed surfaces of freeze fractured samples (Figure 1). No nitrogen or boron spectra were observed for the planar EDS analysis (Supplementary Figures S9 and S10), but a small boron peak and clear nitrogen peak were identified by the analysis software in the cross-sectional EDS from spot acquisitions on the nanoparticles (Supplementary Figures S11 and S12). This confirms that the nanoparticles observed during the atom probe sample preparation were indeed the expected hBN nanoparticles. The lack of boron or nitrogen signal in the planar sample or in the spots adjacent to the nanoparticles in the cross-section sample does not prove that no nanoparticle dissolution has taken place during processing, but it would be extremely challenging to confirm it had occurred using this approach.

Voiding was also observed in regions adjacent to the nanoparticles in the planar freeze fractured samples, but it was not readily apparent if this was a product of the freeze fracture process. Voiding between the polymer bulk and the nanoparticle was observed in the planar SEM images of freeze fractured samples shown in Figure 1. Voiding within nanoparticles themselves was observed in SEM images of FIB prepared cross sections shown in Figure 2 - this additional voiding was not readily observable in the freeze fractured samples due to the roughness of the surface.

The issue of voids causing APT sample preparation failure during the sharpening process was not mitigated to a satisfactory degree and the sample preparation of a bulk polymer/nanoparticle remained challenging. The approach taken was to try and use the simplest methods of sample preparation first, repeated attempts using standard liftout rather than exploration of in-situ void filling and add in further stages as required. The voids are not interconnected and so are only visible during the sharpening stage of sample preparation, meaning that typical bulk methods of filling voids (for example with resin) would not be appropriate as they would not be able to reach the voids (Zand et al., 2023). In situ filling of the nanoscale exposed voids could be carried out through the use of electron beam curing vacuum safe adhesives (Mulcahy et al., 2025) or in-situ sputtering (Woods et al., 2023). However, as these methods require high electron exposure or additional exposure to high energy ions, this could cause additional potential damage to the sample - therefore, they were considered to be less viable than to cross section muliple liftout sample wedges and only attempt to sharpen those with a visible solid nanoparticle/polymer interface. Bulk polymer APT samples were made from those liftout wedges that did not contain a nanoparticle or voiding and were readily sharpened and polished as described in (Thompson et al, 2007).

Previous APT analyses of thin polymer films deposited on metal tips with a diameter of approximately 300 nm (Proudian et al., 2019), have shown to be viable and in fact preferable as the evaporation field of the polymer was low. It should be noted that the evaporation from an insulating solid polymer sample as in this study may not be identical to a thin polymer film on a highly conductive support structure.

A number of bulk polymer samples were successfully fabricated using room temperature Ga FIB, room temperature Xe FIB (Figure 3A and Figure 3B), cryogenic Ga FIB with $D_2$ charging and cryogenic Ga FIB with $D_2$ charging and Cr in-situ coating and analyzed using APT. Only a single atom probe sample containing a visible hBN nanoparticle/bulk polymer interface was successfully fabricated (via room temperature Xe plasma FIB with no deuterium charging) and analyzed by APT (Figure 3b). Both the nanoparticle and the bulk polymer were milled at different rates during the final sharpening process (with the polymer milled more quickly than the nanoparticle), the process ended with a non-optimal apex with a diameter greater than 100 nm and surface roughness visible in the nanoparticle section (Figure 3b).

**Polymer bulk samples**

The major peaks from the mass spectra of the polymer bulk sharpened by room temperature Xe FIB consisted of hydrogen ($H^+$, $H_2^+$, $H_3^+$), carbon ($C^+$, $C^{2+}$) and fragments of the PVDF monomer such as $CF^+$ and $CF_2^+$ (Figure 4). There were no visible peaks associated with boron species, which is expected as the boron nitride particles were unlikely to have been broken up during processing or sample preparation. Peaks at 20 and 21 Da were also observed, which were identified as $FH^+$ and $FH_2^+$ ions from comparison with similar peaks described in the literature (Stoffers et al., 2012). It is possible that the peaks at 20 Da and 21 Da could be an unknown larger compound ion in a higher charge state such as $C_3H_4^{2+}$ and $CH_3H_5^{2+}$, rather than $FH^+$ and $FH_2^+$ or have a mass spectra overlap with such ions. However, these larger compound ions would likely be accompanied by other large carbon ions in lower charge states which are not observed. Quantifying a small overlap between these species would be challenging

(London, 2019) as there may be not enough un-overlapped side peaks that would allow isotopic ratios to be used. After initial fluctuations, the composition of the reconstruction (Figure 5A) and 1D depth concentration profile (Figure 5B) appeared to be uniform with approximate concentrations of ~ 43% at.% C, ~ 30 at.% F, ~ 25 at.% H and ~2 at.% O. The concentration profile data was the decomposed sum of all the peaks shown in Figure 4 and their associated minor isotopes. The exact stoichiometry of the PVDF ($C_2F_2H_2$) was not found, and an excess of carbon was detected.

There are known issues with APT quantification of hydrogen due to hydrogen present in the analysis chamber being detected as part of the material composition, which will increase the apparent concentration of hydrogen as a function of local field and analysis conditions (Sundell et al., 2013). APT analysis of carbon will often under report the amount of carbon due to the multiple hit loss and cluster formation (Thuvander et al., 2019). The number of APT studies that analyse systems containing halides are significantly fewer than for hydrogen or carbon, and so trends of measurement of fluorine may have compositional biases that are not fully understood. The end result is that this is a likely dynamic system where the major elements have artefacts which will affect compositional analysis during the evaporation and detection process and so there is significant scope for fundamental research on how to obtain the known composition of a polymer system.

Similar mass spectra were obtained from samples sharpened with room temperature Ga FIB after $D_2$ charging (Supplementary Figure S13), cryogenic Ga FIB after $D_2$ charging (Supplementary Figure S14) and room temperature Xe plasma FIB (Figure 4). The workflow of cryogenic Ga FIB followed by in-situ Cr coating after $D_2$ charging was not successful in producing a viable dataset. The $D_2$ charging process was not shown to have a visible effect on the sample (Supplementary Figures 15 and 16) and the issue is likely due to the effect of Cr coating on the tip rather than the $D_2$ charging as it was shown to cause bending in the apex

(Supplementary Figure 17). The initial APT data from analyses of the bulk polymer all have the same trend of a transitional region where there are large changes between C, F and H at the apex before moving to a plateau region. The plateau region composition of C, F and H is different between the different workflows, but this is not known whether this is due to the variation in the effect of the workflows as described or if there are unknown variations within them that have not been accounted for (Figure 5, Supplementary Figures S18 and S19). There is a potential overlap with $Ga^+$ and $CF_3^+$ at 69 Da (Figure 4), which means the contribution of that peak to the total carbon content may be affected in those samples sharpened by Ga FIB. The detection of Ga would also indicate potential damage to the sample and so this affect may be hidden due to the overlap.

It is not known exactly in what ionic species that deuterium would be found in the mass spectra of this polymer nanocomposite material, and it is also not known exactly how much deuterium would be introduced into the material and the rate of loss during sample preparation and transfer. Peaks at 1, 2 and 3 Da ($H^+$, $D^+/H_2^+$, $DH^+/H_3^+$), were observed in all uncharged and deuterium charged samples and no peaks at 4 Da ($D_2^+$) were observed in those samples exposed to deuterium gas charging. It is not known exactly in what ionic species that deuterium would be found in the mass spectra of this polymer nanocomposite material, and it is also not known exactly how much deuterium would be introduced into the material and the rate of loss during sample preparation and transfer. The charging parameters for polymer APT samples are not well understood and so there is significant scope for optimisation in terms of charging time, pressure and temperature that could be explored. There were no significant peaks in those compound ion species in the charged samples that would indicate the presence of an $C_nD_m^+$ ion instead of an $C_nH_m^+$ ion in the same manner as OD species have been observed in materials exposed to $D_2$ or $D_2O$ (Martin et al., 2016).

A low concentration single Da shift with the replacement of a hydrogen ion to a deuterium ion into a compound ion would be challenging to quantify, especially given the complexity of the mass spectra. It is likely that an unambiguous deuterium signal would mainly be visible through the $D_2^+$ peak at 4 Da if it was present.

**Sample containing a hBN nanoparticle**

The selection of the polymer matrix and nanoparticles was intended to give distinct visibility between the two phases, however due to both the inherent complexity of polymer mass spectra and related mass spectra overlaps, the identification of all peaks was not possible with certainty. However, there were sufficient differences to be able to distinguish between the mass spectra from bulk polymer and nanoparticles using a number of unambiguous species (Perea et al., 2016) without overlaps such as $B^+$ and $B^{2+}$.

The mass spectrum obtained from the room temperature Xe sharpened sample containing the boron nanoparticle (Figure 6 and Figure 7), there were distinct peaks in single ion species of boron in the $B^+$ (11 Da and 10 Da) and $B^{2+}$ (5.5 Da and 5 Da) charge states (with an approximate ratio of $5B^+:1B^{2+}$) and these were spatially co-segregated with $N^+$ (14 Da) ions in a number of locations within the reconstruction (Figure 8). These can be considered evidence of a boron nitride particle being within the apex of the specimen. The entire dataset was originally assumed to be from a hBN nanoparticle, as was observed from the SEM imaging during sample preparation, however was from a single orientation in the SEM and could be inaccurate. However, they did not form one continuous volume which would be expected from a monolithic hBN nanoparticle - instead they came from multiple regions within the reconstruction. The high contrast nanoparticle at the apex of the Figure 3A may not be physically a single hBN nanoparticle and there may be voids or delamination between layers of hBN that is not visible through SEM imaging. This may explain the dispersed regions high concentration of B and N species within the analysed volume.

The size and distribution could also be consistent with nanoscale sheet separated by bulk polymer or other material than what was considered to be a single, solid particle in the SEM image prior to analysis (Figure 3A). However, as the hBN nanoparticles will have a different evaporation field strength requirement compared to the polymer bulk (likely higher), it is reasonable to assume that some reconstruction artefacts such as compression are present, altering their their apparent size and shape in the reconstruction. There may also have been small microfracture events during analysis, which would potentially match the numerous voltage jumps in the voltage curves (Supplementary Figures S2, S4, S6 and S8) which would have the effect of separating these regions within the reconstruction as the algorithms used to generate the 3D volumes assume a homogeneous and unbroken sample.

The majority of species within the mass spectrum were found in peaks at 20 Da and 21 Da and are identified as $FH^+$ and $FH_2^+$. There were also peaks identified as compound ions of BF that are not expected to exist in the nanoparticle (which should not contain fluorine) or the bulk polymer and peaks that show ions which do not exist directly in that form within the polymer chain, $FH^+$ and $FH^{2+}$ (Figure 7). These are likely to be field-induced compound ions rather than indication of chemical bonds forming between the nanoparticle and the polymer prior to evaporation, especially as there are likely many broken and dangling bonds from the sample preparation.

Multiple minor mass spectra that occur for most integer Da values up to approximately 100 Da can be observed in Figure 6. These are most likely the result of fragments of the polymer chain of varying sizes that have evaporated with a range of hydrogen content. However, it is extremely challenging to reliably determine the identity of each peak. This hydrogen may be supplied by the polymer itself during fragmentation or may be supplied from the chamber itself (Sundell et al. 2013). Although these peaks are small, they have an impact on the ability to quantify the hydrogen content that is present within the polymer sample – hence, the amount of hydrogen that is reported will likely be an underestimate. Despite the relatively large initial diameter of the APT specimen containing the hBN nanoparticle (Figure 3A) compared to that of the bulk polymer sample (Figure 3B), the sample started to evaporate at a relatively low voltage of 2 kV under the analysis conditions used (see Methods). This indicates that these bulk polymers have low evaporation field requirements and may not require to be sharpened to the same dimensions as other materials as indicated by previous work (Proudian et al., 2019).

Elemental B and N species were found to spatially correlate together in three specific regions within the 3D reconstruction (Figure 8). Species identified as the compound ion $BF_2^+$ from their isotopic ratios (Figure 8) were found to be present in the regions above and below those with elemental B and N. This indicates that if they are correctly identified as $BF_2^+$ then the local field conditions may be changing the preferential state of the species being evaporated. However, as they do not spatially segregate with the unambiguous boron and nitrogen species, this seems to be unlikely.

$FH^+$ related species at 20 Da were found to segregate in small clusters to one side of the reconstruction on the -ve X and +ve Y side. This is away from the laser incident side - the laser being incident on the +ve X and -ve Y area of the detector and so is not thought to caused directly by the lasers presence. An example of the detector map of can be found in Supplementary Figure S3, the intensity map shows all ions and increased intensity can be observed in the -ve X and +ve Y area from this species. This was distribution was not observed for the $FH_2^+$ at 21 Da. Small regions of $FH^+$ were also observed at the apex of the bulk polymer fabricated by Xe FIB (Figure S13).

A small amount of Xe ions were found in the top surface regions adjacent to the highest concentration of $B^+$, $B^{2+}$ and $N^+$ species (Figure 9), indicating some residual implantation from the 30 kV sharpening process as it was not possible to use low kV polishing in case the particle was removed from the sample apex. Other bulk polymer samples polished by low kV Xe ions were not found to contain Xe in the surface regions. Despite the lack of Xe detected at the apex of Xe polished samples, this does not mean that the material had not been damaged by knock on damage. The sample may have been damaged by electron imaging or previous ion milling prior to the polishing.

Once in the bulk of the sample, the composition fluctuated by a few percent with the following values of ~ 1% C, ~ 50% F, ~ 35% H, 12% B and ~1% O. stage to an extent that would not have been possible to remove with low kV Xe polishing.

An 2% isoconcentration surface was generated using combined $B^+$ and $B^{2+}$ species to delineate the region associated with the hBN nanoparticle and a cylindrical region of interest was placed over it (Figure 8). Peaks identified as $BF_2^+$ at 48 and 49 Da shown in Figure 7 were not observed to co-segregate with clear $B^+$, $B^{2+}$ or $N^+$ peaks and so their identity as B containing species is not unambiguously confirmed. However, species at 14.5 and 15 Da which are linked to $BF^+$ were observed to co-segregate with the known B and N peaks. This indicates that field variation in the tip in close proximity to the nanoparticle is causing changes in preferential compound ion formation and charge states observed. A dynamic change in mass spectra between dissimilar materials can be expected and to accurately plot the change in composition would need to take this into effect, potentially requiring dynamic peak overlap solving over the interface region (London, 2019).

Oxygen species ($O^+$) at 16 Da were found to co-segregate with B and N species in the region closest to the surface (Figure 8) but not at the other regions of high B and N further down in the reconstruction. This may be linked to Xe ion damage and implantation observed in the surface regions and could be beam related. Species at 28 Da were also found to strongly segregate to the high B and N region at the surface alongside (Figure 7), which could be $N_2^+$. However, a peak at 28 Da was also found in the bulk polymer samples which did not contain any B species, indicating that the ions at 28 Da could be $CO^+$ instead or in addition to $N_2^+$ (Perea et al., 2016). A concentration profile of all potential B containing species was plotted (Figure 10A) along with $N^+$ (14 Da), ions from the peak at 28 Da were not included in the N concentration as they could be either N species or CO species.

Peaks at $B^+$, $B^{2+}$, $BF^+$ and $N^+$ determined to be accurately identified and unambiguous as containing B, as these do not have any identified overlaps and are all observed to co-segregate. $N_2^+$ at 28 Da and $BF_2^+$ at 48 and 49 Da were not shown to be co-segregating with these unambiguous peaks and so were not used in providing a measured ratio of B:N as shown in Figure 10.

The unambiguous peaks of $B^+$, $B_2^+$, $BF^+$ and $N^+$ were then summed to create a 1D concentration profile, which shows an approximate stoichiometry of B:N 2:1 based on peak height (Figure 10B). This almost certainly does not reflect the actual stoichiometry, as there are multiple challenging overlaps that may be impossible to accurately decompose (London, 2019) in addition to known issues of quantifying B (Meisenkothen et al., 2015) and N (Sha et al., 1992).

Hydrogen peaks $H^+$, $H_2^+$ and $H_3^+$ (1 Da, 2 Da and 3 Da) were then plotted along the same concentration profile (Figure 11), showing there is considerable signal from $H^+$ at 1 Da crossing the BN particle along with smaller amounts at 2 Da and 3 Da. This was carried out to determine a baseline level and state of hydrogen that can be expected to be detected between the interface of a hBN nanoparticle and the bulk polymer without any deuterium charging. This indicates that there would be substantial mass spectrum overlaps with peaks of $D^+/H_2^+$ at 2 Da and

$H_3^+$/$DH^+$ at 3 Da and so that it is likely that only a signal at 4 Da from $D_2^+$ would give a reasonable indication of the location of deuterium.

To further investigate the relationship between peaks at 20 and 21 Da labelled as $FH^+$ and $FH_2^+$ and hydrogen species and to determine if there was any dissociation associated with them, multi hit correlative histograms were made of the data collected from the peaks at 20 and 21 labelled as $FH^+$ and $FH_2^+$ and those related to hydrogen species at 1 Da, 2 Da or 3 Da. No indication of co-evaporation behavior or inflight dissociation were found for these species, and the histograms are shown in the Supplementary Information (Supplementary Figures S20-25)

# Discussion

**APT analysis of polymer nanocomposites**

The use of a PVDF matrix and hBN nanoparticles as a model system to analyze polymer nanocomposite materials by APT using site specific FIB liftout was successfully demonstrated, with a variety of FIB workflows facilitating the analysis of nanoscale regions containing both the polymer matrix and the nanocomposite filler. There were significant amounts of voids between the nanoparticles and the polymer matrix within this model system making APT specimen preparation more challenging. In-situ void filling was not attempted for these initial samples in order to reduce potential sample damage and the introduction of initial variables into the worklow. However, for real world applications of such a system, it would be more optimized with respect to nanoparticle aspect ratio, dispersion and distribution to maximize tortuosity and voiding could be substantially reduced or removed. Additionally, surface functionalization of nanoparticles would need to be carried out to improve adhesion to the matrix and reduce voiding, and improve nanoparticle dispersion.

The combined SEM, SEM EDS and APT analyses demonstrated that the regions with high secondary electron contrast as shown in Figure 2 contain unambiguously identified and co-segregated boron and nitrogen species. This indicates that sections of a hBN nanoparticle were present within a prepared APT specimen (Figure 3b) and contributed to the mass spectrum collected during analysis.

The successful detection of specific species that can delineate the polymer matrix (C and F) and the nanoparticles (B and N) within sub volumes of the reconstructed data indicate that the materials selected for the model system were viable. However, some species were identified as containing elements from both the polymer matrix and the nanoparticle ($BF^+$) or having species that were not present in the pristine polymer chain ($FH^+$ and $FH_2^+$). The samples did not have uniform detection maps during analysis (Supplementary Figures S3 and S5) and so reconstruction artefacts are present which would affect both the spatial resolution and the physicality of the reconstructions. Peaks associated with Xe plasma can be observed in the first few nm of the unpolished sample containing the hBN nano particle (Estivill et al., 2016). This indicates that there has been some damage associated with the final FIB milling - however, this would generally be considered as minimal as Xe has a reduced implantation depth compared to other species such as Ga (Eder et al., 2021).

There are known issues which could affect the precise quantification of B (multiple hit loss) (Meisenkothen et al., 2020) and N (dissociation of larger molecules and the formation of neutrals) (Gault et al., 2016) and so variations in these ratios could be expected as shown in previous analysis of hBN nanoparticles in alloy systems (Mohammed et al., 2025), even if there were not issues with mass spectra overlaps. In the analysis of the hBN nanoparticle in this work, there was a combination of mass spectra overlaps and uncertainty of peak identification in addition to the issues raised above and so the ratio of B:N was not found to be 1:1, instead being measured as 2:1.

The composition of the bulk polymers as measured by APT was not consistent during analysis and were extracted from the flattest part of the concentration curves. The plateau concentrations from room temperature Xe FIB (Figure 5B), room temperature Ga FIB sharpening (Supplementary Figure S18) and cryogenic Ga FIB sharpening (Supplementary Figure S19) are shown in Table 1. None of these measured compositions correlate directly with the theoretical monomer compositional ratios of H:F:C 1:1:1. This is not entirely unsurprising given the issues with the quantification of hydrogen species (Chen et al., 2023) due to residual H in the measurement chamber or the formation of neutrals (Gault et al., 2016) and the complex fragmentation and evaporation behavior of carbon (Thuvander et al., 2019) during atom probe analysis.

Oxidation of organic species during SEM has been reported in the literature and oxygen species were also detected in the mass spectra of all samples presented here, in the form of $O^+$ at 16 Da, $OH^+$ at 17 Da and $H_2O^+$ at 18 Da. Water species in those peaks are a known contaminant in APT analyses and the above peaks could be linked to oxidation of the polymer matrix during sample preparation or from contamination. Peaks that could be related to $BO^+$ and $BOH^+$ were also observed in the mass spectra for the particle containing sample (Figure 7), adding further ambiguity into the composition.

While the majority species of the polymer bulk mass spectra could be identified as being carbon and hydrogen related, there were a number of peaks within the nanoparticle sample mass spectra which were unidentified. These may be linked to the polymer matrix evaporating differently in close proximity to the nanoparticle. There is also a possibility that these unknown species are linked to a layer of surface species applied by the manufacturer but not disclosed in provided compositional information. Ionic species that match with m/q values and isotopic ratios matching $BF^+$ compound ions are also present and matched with theoretical isotope ratios, but their prescence would imply that both boron and fluorine species were able to split from their initial bonds and then reform under during mechanical processing or sample preparation. As hBN nanoparticles are extremely stable and fluorine bonds are also stable, it is more likely that the variation in observed compound ions (especially over the region identified as having a high concentration of N as shown in Supplementary Figures S26, S27, S28) are formed via a field induced effect which may have been made more possible by the breaking of bonds by electron or ion beam damage. Analysis of the detector maps show that these samples did not evaporate uniformly, leading to local density variations within the reconstruction – those are especially linked with the high B and N concentration areas linked with the nanoparticles. The density fluctuation is likely linked to the different evaporation field strengths requirements of the two materials, causing local areas of preferential evaporation, especially of the polymer matrix, which can exhibit low evaporation field requirements (Proudian et al., 2019). Microfracture events, which can result in changes in tip apex topography and size, can also result in local variations in surface field. Larger fractures can be observed in the voltage curve by a sudden increase in the required voltage caused by the increase in sample diameter to maintain a set detection rate as can be seen at the end of the voltage profile (as shown in Supplementary Figure S6). Smaller microfractures can be harder to detect as the voltage increases, and can also be similar to those from sudden bursts of evaporation from a region of the surface that requires a lower evaporation field. These density fluctuations will lead to local magnification artefacts in the reconstruction and to a distortion or compressed structure that may not be fully representative of the true composition and structural arrangement (Vurpillot et al., 2013).

**Effect of FIB sample preparation on APT analysis of polymers**

As previously reported in atom probe analyses of polymers, most studies have avoided the requirement for FIB sample preparation and instead through direct deposition onto prefabricated needle shaped specimens. There are a small number of reports of APT specimen preparation using FIB of resin embedded biological material has been previously carried out, with samples analyzed in both laser (Perea et al., 2016) and voltage modes (Adineh et al.,2016) (Adineh et al., 2018) but the effect of beam damage on polymer materials was not fully explored - however, authors do state that bending of sharpened APT specimens under electron or ion imaging had been observed, as was also observed in this study (Supplementary Figure S29). Minimizing electron and ion dose for atom probe analysis has been explored through laser ablation sample preparation for bulk resin-based materials (Sharma et al., 2024). There have also been attempts at fabricating moldable polymer microtips directly onto planar substrates, but these still required FIB milling for the final sharpening (Kostrna et al., 2005). Determining the type and extent of ion beam damage to polymer materials is challenging, especially in-situ on a rapidly changing sample size and shape such as an atom probe specimen during preparation. Analysis of Xe and O plasma FIB cross sectioned polymers using Secondary Electron Hyperspectral Imaging SEM techniques show reduced damage when using O plasma compared to Xe (Farr et al., 2023). However, oxygen plasma has been shown to penetrate more deeply into APT specimens compared to Xe plasma when using direct incidence milling (Eder et al., 2021) and so further experimentation would be required to analyze the effect of oxygen plasma preparation for atom probe specimens with direct incident beams as opposed to cross sections produced with glancing incidence beams. This indicates that there are multiple potential improvements in methodology to ensure that bulk polymers can be prepared for atom probe analysis - however, no rigorous approach exists to determine the optimal route for sample preparation to reduce/eliminate specimen damage.

It is important to note that a superficially viable APT sample fabricated by FIB, which is of appropriate dimensions, appears mechanically stable (even after some beam induced bending) and provides a reasonable mass spectra during analysis does not guarantee that the data is not from a damaged material. Recent rigorous analyses of APT samples subjected to high electron doses during Transmission Kikuchi Diffraction prior to deuterium charging have shown the formation of deuterium trapping defects (Gault et al., 2023). In a number of cases, the use of cryogenic APT sample preparation has shown to cause significant improvements in damage reduction through reducing beam damage in sensitive material systems (Parmenter & Nizamudeen, 2021) (Hayles & De Winter, 2021) and resultant artefact formation (Chang et al., 2019). In this work, there was not a substantial difference in the mass spectra obtained from sharpening samples with cryogenic Ga FIB and room temperature Ga FIB, indicating that the use of cryogenic Ga FIB sharpening may not be sufficient to provide an undamaged sample or that the sample preparation may not be the cause of changes in mass spectra.

Another option would be to reduce or remove the use of FIB sample preparation and instead use existing ultra-microtoming techniques to mechanically remove material from a sharpened tip (Hagler, 2007). Cryogenic TEM of polymers often would use ultramicrotome techniques to be able to reduce damage from FIB and maintain a pristine structure. This could be adapted to prepare either a sharpened polymer tip of a size that is suitable for APT analysis or a partly sharpened tip that could then be sharpened under low dose ion beam conditions to ensure a nanoparticle was contained within the apex.

A known issue of ultramicrotome sample preparation of non-homogeneous materials is the "pop out" of particles with insufficient adhesion to the bulk or with variation in hardness and so this specific model system, with submicron hBN nanoparticles (as in this work) rather than a dispersion of exfoliated hBN nanosheets, may not be as appropriate.

Laser pulsing atom probe analysis is the standard approach for non-conductive materials as voltage pulsing generally leads to sample fracture, however laser pulsing can lead to additional artefacts from thermal effects such as the formation of additional compound ions and non-uniform evaporation related to the laser incident direction. It has been shown that voltage pulsing can be used to analyze non-conductive materials such as resins if a layer of conductive material is applied to the surface of the sharpened atom probe specimen prior to analysis (Adineh et al., 2016). This can be carried out with ex-situ deposition using a sputter coater (Adineh et al., 2016) or in-situ using localized FIB sputtering from a metal source (Schwarz et al., 2024) and may improve the mass spectra quality such that species could be more accurately identified. In-situ coating during cryogenic Ga FIB sharpening was attempted during this study using Cr but no viable bulk polymer data was acquired and sample bending observed. This was likely due to beam damage from the ion beam in close proximity to the sample but room temperature or cryogenic coating from a less damaging method could be attempted using a more distant source (Adineh et al., 2016). This effect also may be due to differences in how the coating adhered and evaporated compared to the polymer and may require optimization of coating material and methodology for polymers.

**Challenges with atom probe mass spectra obtained from polymer nanocomposites**

In addition to the challenges with sample preparation, there are also significant issues with the quantification of carbon species within atom probe (Thuvander et al., 2019) due to multiple hit loss. Longer chain molecules (Meng et al., 2022) have been shown to fragment and dissociate during analysis under high field conditions, with evidence showing preferential cleavage along specific bonds. Compound ions can also be formed during the evaporation from the surface through recombination under high electric field (Brian & Mitchell, 1990) and therefore the existence of a specific compound ion within the mass spectra is not by itself complete evidence of its presence in the pre-evaporated state.

Comparisons were made between mass spectra obtained from polymer bulk samples obtained from Ga FIB, Xe FIB and cryogenic Ga FIB. For each approach, the mass spectra was not appreciably different in terms of the species detected and the fragmentation of the polymer chains. This may indicate that any damage to the material system has not been sufficiently reduced by any of these methods or that the polymer will evaporate in the same manner regardless of the method of FIB sample preparation.

With APT analyses of carbon containing organic systems, the unambiguous identification of the signals in the mass spectra can be challenging as there can be multiple combinations of carbon, nitrogen, oxygen and hydrogen (depending on the material system), effectively giving a peak at every Da value (Figure 6). Due to the high electric field conditions of atom probe analysis, the fragmentation and combination of existing species on the surface prior to evaporation, during evaporation and during flight will affect the mass spectra of the ions detected. Quantifying trends of detected species with sample preparation and analysis parameters would require substantial baseline experiments and even if this was carried out for a number of parameters, each individual atom probe sample will be subtly different in size, shape and composition due to variation in sample preparation and the extremely small volumes involved. This may lead to a situation where dynamic analysis parameter selection in addition to optimized sample preparation, as was determined to be required for gallium nitride systems, (Tang et al., 2015), are required in order to maintain a consistent and accurate composition.

Earlier atom probe studies of polymers where polymers had been deposited into pre-fabricated supports without any associated beam damage showed large fragments of polymer chains – showing fragmentation would be field induced (Maruyama et al., 1987). Atomistic modelling of the evaporation and fragmentation processes of organic molecules from a thin polymer film on a metal surface has been carried out but the behavior of bulk polymer APT samples is likely to be significantly different (Dietrich et al., 2020). For bulk insulating and organic materials, the field penetration depth will include several layers and therefore a larger amount of atoms/bonds of the polymer will experience the electric field. There may also not be a possibility for post ionization as this requires empty electronic states in which electrons can tunnel back into the sample, and it is unclear if this can occur in organic systems.

The lack of larger polymer fragments in this study is likely due to the beam damage induced in the sample preparation, where many bonds have already been broken and so they will more readily evaporate in single ion or small compound ions. The variation in bonding energy between the different polymer chains and different species within individual polymer chains, such as the C-C and the C-C, would also have an impact on the likelihood on their breaking. The lack of difference in mass spectra between samples that were fabricated with different ion species and under different temperatures may imply that the damage reduction approaches were not successful and that more care would be required. However, there is not sufficient understanding of the atom probe analysis of polymers and how they fragment to be able to verify this. It is likely that baseline studies on undamaged polymer samples would be required and therefore would need APT samples to be prepared with no electron or ion beam damage. Unfortunately, the PVDF polymer used in this study is not readily suitable for thin film deposition onto high aspect ratio structures for APT analysis without FIB as it would require melting of the polymer and dipping of the structures, which would be unlikely to form the required dimensions without subsequent FIB milling.

Fluorine compound ions have been previously observed in the literature when analyzing fluorinated self-assembled monolayers (SAMs) (as organic thin films) with APT (Stoffers et al., 2012). That work showed that there can be links with the layering and orientation of specific functional groups with respect to the substrate in the case of gold coated with

(1H,1H,2H,2H-)perfluoro-decanethiol (PFDT) SAMs, where most of hydrogen molecules are replaced by much heavier fluorine. The authors were able to show a field dependence in the evaporation sequence that can be linked to the severing of specific bonds within the molecule, and such a system may be suitable for determining the extent at which electron or ion beam damage is present by the change in evaporation behavior from a beam exposed sample and non beam exposed sample. In other works, fluorine containing compound ions, including $CF_3^+$, $FH^+$ and $FH_2^+$ have been observed in atom probe analysis (Chattopadhyay, 2010).

In the bulk polymer samples, the mass spectra was found to be mainly carbon $C^+$ and $C_2^+$ species with some $C_xF_y$ species and some unidentified peaks likely to be $C_nH_m$ species - this is consistent with what was previously found in the literature for APT analysis of fluorinated hydrocarbons and organic materials. The mass spectra did not show evidence of longer sections of the polymer chains in the form of multiple connected monomers. However, in this case, where the field conditions preferentially evaporate smaller molecular ions and single ions, the small datasets (with the largest one at the order of a million of ions) may not be large enough to detect larger complex ions in low concentrations.

In the sample containing a hBN nanoparticle, a number of co-segregating regions of elemental B and N were observed which could be linked to single sheets of hBN. However, the entire sample did not consist of hBN as would be expected if the sample apex was indeed a monolithic nanoparticle (Figure 8). Species identified as $BF_2^+$ were present above and below these regions, but additional nitrogen species were not identified outside of these regions. This sample contained $FH^+$ and $FH_2^+$ species that made up the majority of the ions in the sample, giving a non-stoichiometric excess of F if we were to assume that any region not consisting of hBN should follow the approximate composition of PVDF. These regions of $FH^+$ and $FH_2^+$ were non-uniform within the reconstruction and $FH^+$ was found to appear as point clusters, which is also unexpected if the material was uniform. This implies that the tip may have experienced non-uniform evaporation behavior due to field differences across the tip with relation to the laser, with preferential evaporation of $FH^+$ and $FH_2^+$ species over carbon species.

For future studies, it is important to understand how organic/carbon-based materials fragment and field evaporate, to improve the identification of species and give some scope as to the potential information that can be reliably obtained by APT analysis. There is a lack of theoretical understanding of how these polymer systems evaporate under high fields, especially in bulk rather than thin films, where the electric field is concentrated at the tip apex.

**Deuterium charging of polymer samples**

As it was not possible to obtain data from a $D_2$ charged sample which contained a hBN nanoparticle/bulk polymer interface due to sample preparation challenges, a more full understanding of the deuterium trapping behaviour in this polymer system as not possible. There were significant peaks at 1, 2 and 3 Da in samples from both bulk polymer (Supplementary Figure S14) and from those containing a hBN nanoparticle (Figure 7) - they were assigned as $H^+$, $H_2^+$ and $H_2^{+\cdot}$. The H signal can originate either from the polymer matrix or from the measurement - however, the use of deuterium is required for charging experiments rather than hydrogen due to the mass spectra overlapping in those peaks. Compound deuterium at $D_2^+$ at 4 Da is an unambiguous signal of deuterium being present (Gault et al., 2024) but the proportion of the deuterium observed in that species is generally low and so there can be issues with low concentration segregation. No visible peaks at 4 Da were observed in charged samples within the experiments carried out in this work - but as nanoparticles are expected to be trapping sites for hydrogen within the system and as no nanoparticles were successfully placed at the apex of a sample prior to charging, it is perhaps unsurprising that deuterium was not observed. It is also unlikely that deuterium would be able to replace a hydrogen atom on the polymer chain during the charging condition used and so any deuterium segregation would likely be interstitial rather than substitutional and so would quickly diffuse out, again reducing the chances of detection in the polymer bulk. It is unknown how the deuterium would permeate or segregate within a potentially damaged polymer system compared to a pristine system and so the post charging cooling time may not have been quick enough to prevent the escape of any deuterium.

## Conclusions

Site specific atom probe sample preparation by FIB liftout of bulk polymer nanocomposites has been shown for the first time in a model system of hexagonal boron nitride (hBN) nanoparticles within a PVDF polymer matrix. The nanoparticles were sufficiently large and had sufficient electron imaging contrast that they could be readily observed within the SEM during sample preparation – they could be placed within the apex of an atom probe specimen, albeit with challenges due to voiding causing structural instability. A variety of sample preparation pathways were carried out in an effort to find if there were any obvious variations in sample quality, yield and mass spectra. These variations included room temperature Ga FIB liftout and sharpening, room temperature Xe FIB liftout and sharpening, cryogenic Ga FIB sharpening and in-situ cryogenic Cr coating and deuterium ($D_2$) gas charging.

Mass spectra from the bulk polymer and the nanoparticle were collected using pulsed laser atom probe using standard conditions and compared. Only one sample containing a nanoparticle/bulk polymer interface was successfully analyzed due to delamination associated with voiding at the polymer bulk/nanoparticle interface which caused APT specimens to fail during the final sharpening process. This sample was not charged with $D_2$ and only bulk polymer samples sharpened by cryogenic Ga FIB were successfully charged and analysed by APT.

Bulk polymer samples showed mass spectra associated with fragments and ions from the polymer chain in the form of $C^+$, $C_2^+$ and $C_nH_m$ species but no complete monomers. The sample containing hBN nanoparticle showed regions of co-segregating B and N and significant amounts of $FH^+$ and $FH_2^+$ species – this gave an excess of F species within the reconstruction, as would be expected from PDVF stoichiometry. No peaks associated with deuterium ($D_2$) were observed in the charged PVDF samples – further work is needed to look into gas charged hBN nanoparticles and hBN/PVDF systems with various hBN vacancies and nanocomposite interfaces expected to be trapping sites for hydrogen/deuterium.

The main issues with site specific FIB liftout of polymer nanocomposites appears to be the damage to the polymer matrix (PVDF) from the sample preparation and the unknown field evaporation behavior of bulk polymers, both of which cause mass spectrum challenges in the form of unknown peaks and non-consistent polymer compositions. This implies that quantitative analysis of polymer systems requires further development. Additionally, as the hBN nanoparticles were more easily observed, these current workflows may be more appropriate for investigations where the nanoparticles are the main focus rather than the interaction between the nanoparticles and polymer matrix. It was also challenging to determine at which stage and the extent of beam damage was experienced by the samples in the different preparation workflows. However, as they all consisted of room temperature FIB liftout protocols prior to the different sharpening and charging processes, a fully cryogenic pathway during liftout may be required to reduce potential damage.

In summary, this work has shown the initial development and challenges associated with site specific FIB sample preparation and APT analysis for polymer nanocomposites, opening up new options for nanoscale analysis in these previously unexplored systems.

**Acknowledgements**

JD, SP and BG are grateful for funding from the EPSRC under the grant EP/V007661/1.

AS acknowledges the financial support from Deutsche Forschungsgemeinschaft (DFG) under project A4 of the collaborative research center SFB/TR 103.

TMS. gratefully acknowledges the financial support of the Walter Benjamin Program of the German Research Foundation (DFG) (Project No. 551061178). T.M. S. and B.G. are grateful for funding from the DFG through the award of the Leibniz Prize 2020 from B.G. Uwe Tezins, Andreas Sturm and Christian Broß are acknowledged for their support to the FIB & APT facilities at MPI-SusMat.

LF acknowledges financial support through the Research Development Fund (RDF) (RD22106).

EMP acknowledges financial support from UKRI's Future Leaders Fellowship programme [grant MR/V024124/1] and from the UKRI Horizon Europe Guarantee programme (ERC Starting Grant *ResistHfracture*, EP/Y037219/1).

SP is funded by EPSRC fellowship EP/SO13881/1 and an RAEng Associate research fellowship.

**References**

Adineh, V. R., Zheng, C., Zhang, Q., Marceau, R. K. W., Liu, B., Chen, Y., Si, K. J., Weyland, M., Velkov, T., Cheng, W., Li, J., & Fu, J. (2018). Graphene-Enhanced 3D Chemical Mapping of Biological Specimens at Near-Atomic Resolution. *Advanced Functional Materials*, *28*(32), 1801439. https://doi.org/https://doi.org/10.1002/adfm.201801439

Adineh, V. R., Marceau RKW, Velkov T, Li J. & Fu J. (2016). Near-atomic three-dimensional mapping for site-specific chemistry of 'superbugs'. *Nano Lett 16*, 7113–7120. 10.1021/acs.nanolett.6b03409

Bagot, P. A. J., Silk, O. B. W., Douglas, J. O., Pedrazzini, S., Crudden, D. J., Martin, T. L., Hardy, M. C., Moody, M. P., & Reed, R. C. (2017). An Atom Probe Tomography study of site preference and partitioning in a nickel-based superalloy. *Acta Materialia*, *125*, 156–165. https://doi.org/https://doi.org/10.1016/j.actamat.2016.11.053

Bassm, N. D., De Gregorio, B. T., Kilcoyne, A. L. D., Scott, K., Chou, T., Wirick, S., Cody, G. and Stroud, R.M. (2012), Minimizing damage during FIB sample preparation of soft materials. Journal of Microscopy, 245: 288-301. https://doi.org/10.1111/j.1365-2818.2011.03570.x

Brian, J., & Mitchell, A. (1990). The dissociative recombination of molecular ions. *Physics Reports*, *186*(5), 215–248. https://doi.org/https://doi.org/10.1016/0370-1573(90)90159-Y

Bunton, J. H., Olson, J. D., Lenz, D. R., & Kelly, T. F. (2007). Advances in Pulsed-Laser Atom Probe: Instrument and Specimen Design for Optimum Performance. *Microscopy and Microanalysis*, *13*(6), 418–427. https://doi.org/10.1017/S1431927607070869

Chang, Y., Lu, W., Guénolé, J., Stephenson, L. T., Szczpaniak, A., Kontis, P., Ackerman, A. K., Dear, F. F., Mouton, I., Zhong, X., Zhang, S., Dye, D., Liebscher, C. H., Ponge, D., Korte-Kerzel, S., Raabe, D., & Gault, B. (2019). Ti and its alloys as examples of cryogenic focused ion beam milling of environmentally-sensitive materials. *Nature Communications*, *10*(1), 942. https://doi.org/10.1038/s41467-019-08752-7

Chattopadhyay, A. (2010). A comparative study of the spectroscopic features of the low-lying electronic states of H2F+ and H2Cl+ ions. *Journal of Chemical Sciences*, *122*(2), 259–269. https://doi.org/10.1007/s12039-010-0030-y

Chen, Y.-S., Haley, D., Gerstl, S. S. A., London, A. J., Sweeney, F., Wepf, R. A., Rainforth, W. M., Bagot, P. A. J., & Moody, M. P. (2017). Direct observation of individual hydrogen atoms at trapping sites in a ferritic steel. *Science*, *355*(6330), 1196–1199. https://doi.org/10.1126/science.aal2418

Chen, Y.-S., Liu, P.-Y., Niu, R., Devaraj, A., Yen, H.-W., Marceau, R. K. W., & Cairney, J. M. (2023). Atom Probe Tomography for the Observation of Hydrogen in Materials: A Review. *Microscopy and Microanalysis*, *29*(1), 1–15. https://doi.org/10.1093/micmic/ozac005

Douglas, J.O., Conroy, M., Giuliani, F., Gault, B. (2023) *In Situ* Sputtering From the Micromanipulator to Enable Cryogenic Preparation of Specimens for Atom Probe Tomography by Focused-Ion Beam, *Microscopy and Microanalysis*, Volume 29, Issue 3, June 2023, Pages 1009–1017, https://doi.org/10.1093/micmic/ozad020

Daly, L., Lee, M. R., Hallis, L. J., Ishii, H. A., Bradley, J. P., Bland, Phillip. A., Saxey, D. W., Fougerouse, D., Rickard, W. D. A., Forman, L. V, Timms, N. E., Jourdan, F., Reddy, S. M., Salge, T., Quadir, Z., Christou, E., Cox, M. A., Aguiar, J. A., Hattar, K., … Thompson, M. S. (2021). Solar wind contributions to Earth's oceans. *Nature Astronomy*, *5*(12), 1275–1285. https://doi.org/10.1038/s41550-021-01487-w

Dietrich, C. A., Schuldt, R., Born, D., Solodenko, H., Schmitz, G., & Kästner, J. (2020). Evaporation and Fragmentation of Organic Molecules in Strong Electric Fields Simulated with DFT. *The Journal of Physical Chemistry A*, *124*(41), 8633–8642. https://doi.org/10.1021/acs.jpca.0c06887

Eder, K., Bhatia, V., Qu, J., Van Leer, B., Dutka, M., & Cairney, J. M. (2021). A multi-ion plasma FIB study: Determining ion implantation depths of Xe, N, O and Ar in tungsten via atom probe tomography. *Ultramicroscopy*, *228*, 113334. https://doi.org/https://doi.org/10.1016/j.ultramic.2021.113334

Ellingford, C., Wan, C., Figiel, L., McNally T. (2018). Mechanical and dielectric properties of MWCNT filled chemically modified SBS/PVDF blends. *Composites Communications*, 8, 58-64. https://doi.org/10.1016/j.coco.2017.11.001

Estivill, R., Audoit, G., Barnes, J.-P., Grenier, A., & Blavette, D. (2016). Preparation and Analysis of Atom Probe Tips by Xenon Focused Ion Beam Milling. *Microscopy and Microanalysis*, *22*(3), 576–582. https://doi.org/10.1017/S1431927616000581

Farr, N. T. H., Pasniewski, M., & de Marco, A. (2023). Assessing the Quality of Oxygen Plasma Focused Ion Beam (O-PFIB) Etching on Polypropylene Surfaces Using Secondary Electron Hyperspectral Imaging. *Polymers*, *15*(15). https://doi.org/10.3390/polym15153247

Figiel, L., De Angelis, M.G., Janssen, F., Vehlow, D., Giannis, S., Skytree, L., Walls-Bruck, M., & Douglas, A. (2025). Polymers and Composites for Hydrogen Economy. *Journal of Materials Science: Composites*, 6(11). https://doi.org/10.1186/s42252-025-00076-8

Gajjela, R. S. R., Hendriks, A. L., Douglas, J. O., Sala, E. M., Steindl, P., Klenovský, P., Bagot, P. A. J., Moody, M. P., Bimberg, D., & Koenraad, P. M. (2021). Structural and compositional analysis of (InGa)(AsSb)/GaAs/GaP Stranski–Krastanov quantum dots. *Light: Science & Applications*, *10*(1), 125. https://doi.org/10.1038/s41377-021-00564-z

Gault, B., Chiaramonti, A., Cojocaru-Mirédin, O., Stender, P., Dubosq, R., Freysoldt, C., Makineni, S. K., Li, T., Moody, M., & Cairney, J. M. (2021). Atom probe tomography. *Nature Reviews Methods Primers*, *1*(1), 51. https://doi.org/10.1038/s43586-021-00047-w

Gault, B., Khanchandani, H., Prithiv, T. S., Antonov, S., & Britton, T. Ben. (2023). Transmission Kikuchi Diffraction Mapping Induces Structural Damage in Atom Probe Specimens. *Microscopy and Microanalysis*, *29*(3), 1026–1036. https://doi.org/10.1093/micmic/ozad029

Gault, B., Saksena, A., Sauvage, X., Bagot, P., Aota, L. S., Arlt, J., Belkacemi, L. T., Boll, T., Chen, Y.-S., Daly, L., Djukic, M. B., Douglas, J. O., Duarte, M. J., Felfer, P. J., Forbes, R. G., Fu, J., Gardner, H. M., Gemma, R., Gerstl, S. S. A., … Zou, B. (2024). Towards Establishing Best Practice in the Analysis of Hydrogen and Deuterium by Atom Probe Tomography. *Microscopy and Microanalysis*, *30*(6), 1205–1220. https://doi.org/10.1093/mam/ozae081

Gault, B., Saxey, D. W., Ashton, M. W., Sinnott, S. B., Chiaramonti, A. N., Moody, M. P., & Schreiber, D. K. (2016). Behavior of molecules and molecular ions near a field emitter. *New Journal of Physics*, *18*(3), 33031. https://doi.org/10.1088/1367-2630/18/3/033031

Giddings, A. D., Koelling, S., Shimizu, Y., Estivill, R., Inoue, K., Vandervorst, W., & Yeoh, W. K. (2018). Industrial application of atom probe tomography to semiconductor devices. *Scripta Materialia*, *148*, 82–90. https://doi.org/https://doi.org/10.1016/j.scriptamat.2017.09.004

Grandfield, K., Micheletti, C., Deering, J., Arcuri, G., Tang, T., & Langelier, B. (2022). Atom probe tomography for biomaterials and biomineralization. *Acta Biomaterialia*, *148*, 44–60. https://doi.org/https://doi.org/10.1016/j.actbio.2022.06.010

Hagler, H. K. (2007). Ultramicrotomy for Biological Electron Microscopy. In J. Kuo (Ed.), *Electron Microscopy: Methods and Protocols* (pp. 67–96). Humana Press. https://doi.org/10.1007/978-1-59745-294-6_5

Hayles, M. F., & De Winter, D. A. M. (2021). An introduction to cryo-FIB-SEM cross-sectioning of frozen, hydrated Life Science samples. *Journal of Microscopy*, *281*(2), 138–156. https://doi.org/https://doi.org/10.1111/jmi.12951

Herbig, M. (2018). Spatially correlated electron microscopy and atom probe tomography: Current possibilities and future perspectives. *Scripta Materialia*, *148*, 98–105. https://doi.org/https://doi.org/10.1016/j.scriptamat.2017.03.017

Herbig, M., Choi, P., & Raabe, D. (2015). Combining structural and chemical information at the nanometer scale by correlative transmission electron microscopy and atom probe tomography. *Ultramicroscopy*, *153*, 32–39. https://doi.org/https://doi.org/10.1016/j.ultramic.2015.02.003

Jayaram, R., Russell, K. F., & Miller, M. K. (1993). An atom probe study of the substitutional behavior of beryllium in NiAl. *Applied Surface Science*, *67*(1), 316–320. https://doi.org/https://doi.org/10.1016/0169-4332(93)90332-6

Jenkins, B. M., Douglas, J. O., Gardner, H. M., Tweddle, D., Kareer, A., Karamched, P. S., Riddle, N., Hyde, J. M., Bagot, P. A. J., Odette, G. R., & Moody, M. P. (2020). A more holistic characterisation of internal interfaces in a variety of materials via complementary use of transmission Kikuchi diffraction and Atom probe tomography. *Applied Surface Science*, *528*, 147011. https://doi.org/https://doi.org/10.1016/j.apsusc.2020.147011

Jung, J. K., Kim, I. G., Chung, K. S., & Baek, U. B. (2021). Gas chromatography techniques to evaluate the hydrogen permeation characteristics in rubber: ethylene propylene diene monomer. *Scientific Reports*, *11*(1), 4859. https://doi.org/10.1038/s41598-021-83692-1

Kelly, T. F., & Miller, M. K. (2007). Atom probe tomography. *Review of Scientific Instruments*, *78*(3), 31101. https://doi.org/10.1063/1.2709758

Kostrna, S. L. P., Mengelt, T. J., Ali, M., Larson, D. J., Kelly, T. F., & Goodman, S. L. (2005). Creating Polymer Microtip Specimens for Atom Probe Tomography. *Microscopy and Microanalysis*, *11*(S02), 874–875. https://doi.org/10.1017/S143192760551047X

Larson, D. J., Giddings, A. D., Wu, Y., Verheijen, M. A., Prosa, T. J., Roozeboom, F., Rice, K. P., Kessels, W. M. M., Geiser, B. P., & Kelly, T. F. (2015). Encapsulation method for atom probe tomography analysis of nanoparticles. *Ultramicroscopy*, *159*, 420–426. https://doi.org/https://doi.org/10.1016/j.ultramic.2015.02.014

Liu, M., Lin, K., Zhou, M., Wallwork, A., Bissett, M. A., Young, R. J., & Kinloch, I. A. (2024). Mechanism of gas barrier improvement of graphene/polypropylene nanocomposites for new-generation light-weight hydrogen storage. *Composites Science and Technology*, *249*, 110483. https://doi.org/https://doi.org/10.1016/j.compscitech.2024.110483

London, A. J. (2019). Quantifying Uncertainty from Mass-Peak Overlaps in Atom Probe Microscopy. *Microscopy and Microanalysis*, *25*(2), 378–388. https://doi.org/10.1017/S1431927618016276

Marquis, E. A., Geiser, B. P., Prosa, T. J., & Larson, D. J. (2011). Evolution of tip shape during field evaporation of complex multilayer structures. *Journal of Microscopy*, *241*(3), 225–233. https://doi.org/https://doi.org/10.1111/j.1365-2818.2010.03421.x

Martin, T. L., Coe, C., Bagot, P. A. J., Morrall, P., Smith, G. D. W., Scott, T., & Moody, M. P. (2016). Atomic-scale Studies of Uranium Oxidation and Corrosion by Water Vapour. *Scientific Reports*, *6*(1), 25618. https://doi.org/10.1038/srep2561

Maruyama, T., Hasegawa, Y., Nishi, T., & Sakurai, T. (1987). FIM And Atom-Probe Study of Polymers. *J. Phys. Colloques*, *48*(C6), C6-274. https://doi.org/10.1051/jphyscol:1987644

McGilvery, C. M., Abellan, P., Kłosowski, M. M., Livingston, A. G., Cabral, J. T., Ramasse, Q. M., & Porter, A. E. (2020). Nanoscale Chemical Heterogeneity in Aromatic Polyamide Membranes for Reverse Osmosis Applications. *ACS Applied Materials & Interfaces*, *12*(17), 19890–19902. https://doi.org/10.1021/acsami.0c01473

Meisenkothen, F., Samarov, D. V, Kalish, I., & Steel, E. B. (2020). Exploring the accuracy of isotopic analyses in atom probe mass spectrometry. *Ultramicroscopy*, *216*, 113018. https://doi.org/https://doi.org/10.1016/j.ultramic.2020.113018

Meisenkothen, F., Steel, E. B., Prosa, T. J., Henry, K. T., & Prakash Kolli, R. (2015). Effects of detector dead-time on quantitative analyses involving boron and multi-hit detection events in atom probe tomography. *Ultramicroscopy*, *159*, 101–111. https://doi.org/https://doi.org/10.1016/j.ultramic.2015.07.009

Meng, K., Schwarz, T. M., Weikum, E. M., Stender, P., & Schmitz, G. (2022). Frozen n-Tetradecane Investigated by Cryo-Atom Probe Tomography. *Microscopy and Microanalysis*, *28*(4), 1289–1299. https://doi.org/10.1017/S143192762101254X

Mohammed, S. M. A. K., Li, Z., Orikasa, K., Devaraj, A., Garcia, D., Sarvesha, R., Lama, A., Pole, M., Park, C., Chu, S.-H., Ross, K., & Agarwal, A. (2025). Neutron radiation induced transmutation of boron to lithium in aluminum-boron nitride composite. *Materials Today Advances*, *25*, 100551. https://doi.org/10.1016/j.mtadv.2024.100551

Mu, M., Teblum, E., Figiel, Ł., Nessim, G.D., McNally, T. (2018). Correlation between MWCNT aspect ratio and the mechanical properties of composites of PMMA and MWCNTs. *Materials Research Express*, 5(4), 045305. https://doi.org/10.1088/2053-1591/aab82d Mulcahy, N., Douglas, J. O., & Conroy, M. S. (2025). Look What You Made Me Glue: SEMGluTM Enabled Alternative Cryogenic Sample Preparation Process for Cryogenic Atom Probe Tomography Studies. *Microscopy and Microanalysis*, *31*(4), ozaf063. https://doi.org/10.1093/mam/ozaf063

Nakagawa, K., & Ishida, Y. (1973). Annealing effects in poly(vinylidene fluoride) as revealed by specific volume measurements, differential scanning calorimetry, and electron microscopy. *Journal of Polymer Science: Polymer Physics Edition*, *11*(11), 2153–2171. https://doi.org/https://doi.org/10.1002/pol.1973.180111107

Nishikawa, O., & Kato, H. (1986). Atom-probe study of a conducting polymer: The oxidation of polypyrrole. *The Journal of Chemical Physics*, *85*(11), 6758–6764. https://doi.org/10.1063/1.451407

Parmenter, C. D., & Nizamudeen, Z. A. (2021). Cryo-FIB-lift-out: practically impossible to practical reality. *Journal of Microscopy*, *281*(2), 157–174. https://doi.org/https://doi.org/10.1111/jmi.12953

Perea, D. E., Liu, J., Bartrand, J., Dicken, Q., Thevuthasan, S. T., Browning, N. D., & Evans, J. E. (2016). Atom Probe Tomographic Mapping Directly Reveals the Atomic Distribution of Phosphorus in Resin Embedded Ferritin. *Scientific Reports*, *6*(1), 22321. https://doi.org/10.1038/srep22321

Prosa, T. J., Keeney Y, S. K., & Kelly, T. F. (2010). Atom probe tomography analysis of poly(3-alkylthiophene)s. *Journal of Microscopy*, *237*(2), 155–167. https://doi.org/https://doi.org/10.1111/j.1365-2818.2009.03320.x

Proudian, A. P., Jaskot, M. B., Diercks, D. R., Gorman, B. P., & Zimmerman, J. D. (2019). Atom Probe Tomography of Molecular Organic Materials: Sub-Dalton Nanometer-Scale Quantification. *Chemistry of Materials*, *31*(7), 2241–2247. https://doi.org/10.1021/acs.chemmater.8b04476

Quaresimin, M., Schulte, K., Zappalorto, M., & Chandrasekaran, S. (2016). Toughening mechanisms in polymer nanocomposites: From experiments to modelling. *Composites Science and Technology*, *123*, 187–204. https://doi.org/https://doi.org/10.1016/j.compscitech.2015.11.027

Reddy, S. M., Saxey, D. W., Rickard, W. D. A., Fougerouse, D., Montalvo, S. D., Verberne, R., & van Riessen, A. (2020). Atom Probe Tomography: Development and Application to the Geosciences. *Geostandards and Geoanalytical Research*, *44*(1), 5–50. https://doi.org/https://doi.org/10.1111/ggr.12313

Saksena, A., Sun, B., Dong, X., Khanchandani, H., Ponge, D., & Gault, B. (2024). Optimizing site-specific specimen preparation for atom probe tomography by using hydrogen for visualizing radiation-induced damage. *International Journal of Hydrogen Energy*, *50*, 165–174. https://doi.org/https://doi.org/10.1016/j.ijhydene.2023.09.057

Schreiber, D. K., Perea, D. E., Ryan, J. V, Evans, J. E., & Vienna, J. D. (2018). A method for site-specific and cryogenic specimen fabrication of liquid/solid interfaces for atom probe tomography. *Ultramicroscopy*, *194*, 89–99. https://doi.org/https://doi.org/10.1016/j.ultramic.2018.07.010

Schwarz, T. M., Woods, E., Singh, M. P., Chen, X., Jung, C., Aota, L. S., Jang, K., Krämer, M., Kim, S-H., McCarroll, I. & Gault, B. (2024). In situ metallic coating of atom probe specimen for enhanced yield, performance, and increased field-of-view. *Microsc Microanal ozae006*. https://academic.oup.com/mam/advance-article/doi/10.1093/mam/ozae006/7608750

Sha, W., Chang, L., Smith, G. D. W., Cheng, L., & Mittemeijer, E. J. (1992). Some aspects of atom-probe analysis of Fe C and Fe N systems. *Surface Science*, *266*(1), 416–423. https://doi.org/https://doi.org/10.1016/0039-6028(92)91055-G

Sharma, A., Tegg, L., Djoulde, A., Marla, D., & s, J. (2024). Rapid Preparation of Nanoscale Resin-Embedded Samples Using Site-Specific Laser Ablation and Focused Ion Beam Milling. *Microscopy and Microanalysis*, *30*(Supplement_1), ozae044.046-ozae044.046. https://doi.org/10.1093/mam/ozae044.046

Smith, A., Kelly N.L., Figiel, Ł., Wan C., Hanna, J.V., Farris, S., McNally T. (2020). Graphene Oxide Functionalized with 2-Ureido-4 [1H]-pyrimidinone for production of nacre-like films. *ACS Applied Nano Materials* 3(7) 7161-7171, https://doi.org/10.1021/acsanm.0c01488

Stephenson, L. T., Szczepaniak, A., Mouton, I., Rusitzka, K. A. K., Breen, A. J., Tezins, U., Sturm, A., Vogel, D., Chang, Y., Kontis, P., Rosenthal, A., Shepard, J. D., Maier, U., Kelly, T. F., Raabe, D., & Gault, B. (2018). The Laplace Project: An integrated suite for preparing and transferring atom probe samples under cryogenic and UHV conditions. *PLOS ONE*, *13*(12), e0209211. https://doi.org/10.1371/journal.pone.0209211

Stoffers, A., Oberdorfer, C., & Schmitz, G. (2012). Controlled Field Evaporation of Fluorinated Self-Assembled Monolayers. *Langmuir*, *28*(1), 56–59. https://doi.org/10.1021/la204126x

Sundell, G., Thuvander, M., & Andrén, H.-O. (2013). Hydrogen analysis in APT: Methods to control adsorption and dissociation of H2. *Ultramicroscopy*, *132*, 285–289. https://doi.org/https://doi.org/10.1016/j.ultramic.2013.01.007

Takahashi, J., Kawakami, K., Kobayashi, Y., & Tarui, T. (2010). The first direct observation of hydrogen trapping sites in TiC precipitation-hardening steel through atom probe tomography. *Scripta Materialia*, *63*(3), 261–264. https://doi.org/https://doi.org/10.1016/j.scriptamat.2010.03.012

Thompson, K., Lawrence, D., Larson, D. J., Olson, J. D., Kelly, T. F., & Gorman, B. (2007). In situ site-specific specimen preparation for atom probe tomography. *Ultramicroscopy*, *107*(2), 131–139. https://doi.org/https://doi.org/10.1016/j.ultramic.2006.06.008

Thuvander, M., Shinde, D., Rehan, A., Ejnermark, S., & Stiller, K. (2019). Improving Compositional Accuracy in APT Analysis of Carbides Using a Decreased Detection Efficiency. *Microscopy and Microanalysis*, *25*(2), 454–461. https://doi.org/10.1017/S1431927619000424

Tweddle, D., Hamer, P., Shen, Z., Markevich, V. P., Moody, M. P., & Wilshaw, P. R. (2021). Direct observation of hydrogen at defects in multicrystalline silicon. *Progress in Photovoltaics: Research and Applications*, *29*(11), 1158–1164. https://doi.org/https://doi.org/10.1002/pip.3184

Unalan, I. U., Wan, C., Figiel, Ł., Olsson, R. T., Trabattoni, S., & Farris, S. (2015). Exceptional oxygen barrier performance of pullulan nanocomposites with ultra-low loading of graphene oxide. *Nanotechnology*, *26*(27), 275703. https://doi.org/10.1088/0957-4484/26/27/275703

Vurpillot, F., Gault, B., Geiser, B. P., & Larson, D. J. (2013). Reconstructing atom probe data: A review. *Ultramicroscopy*, *132*, 19–30. https://doi.org/https://doi.org/10.1016/j.ultramic.2013.03.010

Woods, E. V, Singh, M. P., Kim, S.-H., Schwarz, T. M., Douglas, J. O., El-Zoka, A. A., Giulani, F., & Gault, B. (2023). A Versatile and Reproducible Cryo-sample Preparation Methodology for Atom Probe Studies. *Microscopy and Microanalysis*, *29*(6), 1992–2003. https://doi.org/10.1093/micmic/ozad120

Yang, Q., Danaie, M., Young, N., Broadley, V., Joyce, D. E., Martin, T. L., Marceau, E., Moody, M. P., & Bagot, P. A. J. (2019). Atom Probe Tomography of Au–Cu Bimetallic Nanoparticles Synthesized by Inert Gas Condensation. *The Journal of Physical Chemistry C*, *123*(43), 26481–26489. https://doi.org/10.1021/acs.jpcc.9b09340

Zand, F., Hangx, S. J. T., Spiers, C. J., van den Brink, P. J., Burns, J., Boebinger, M. G., Poplawsky, J. D., Monai, M., & Weckhuysen, B. M. (2023). Elucidating the Structure and Composition of Individual Bimetallic Nanoparticles in Supported Catalysts by Atom Probe Tomography. *Journal of the American Chemical Society*, *145*(31), 17299–17308. https://doi.org/10.1021/jacs.3c04474

Zanuttini, D., Blum, I., Rigutti, L., Vurpillot, F., Douady, J., Jacquet, E., Anglade, P.-M., & Gervais, B. (2017). Simulation of field-induced molecular dissociation in atom-probe tomography: Identification of a neutral emission channel. *Physical Review A*, *95*(6), 61401. https://doi.org/10.1103/PhysRevA.95.061401

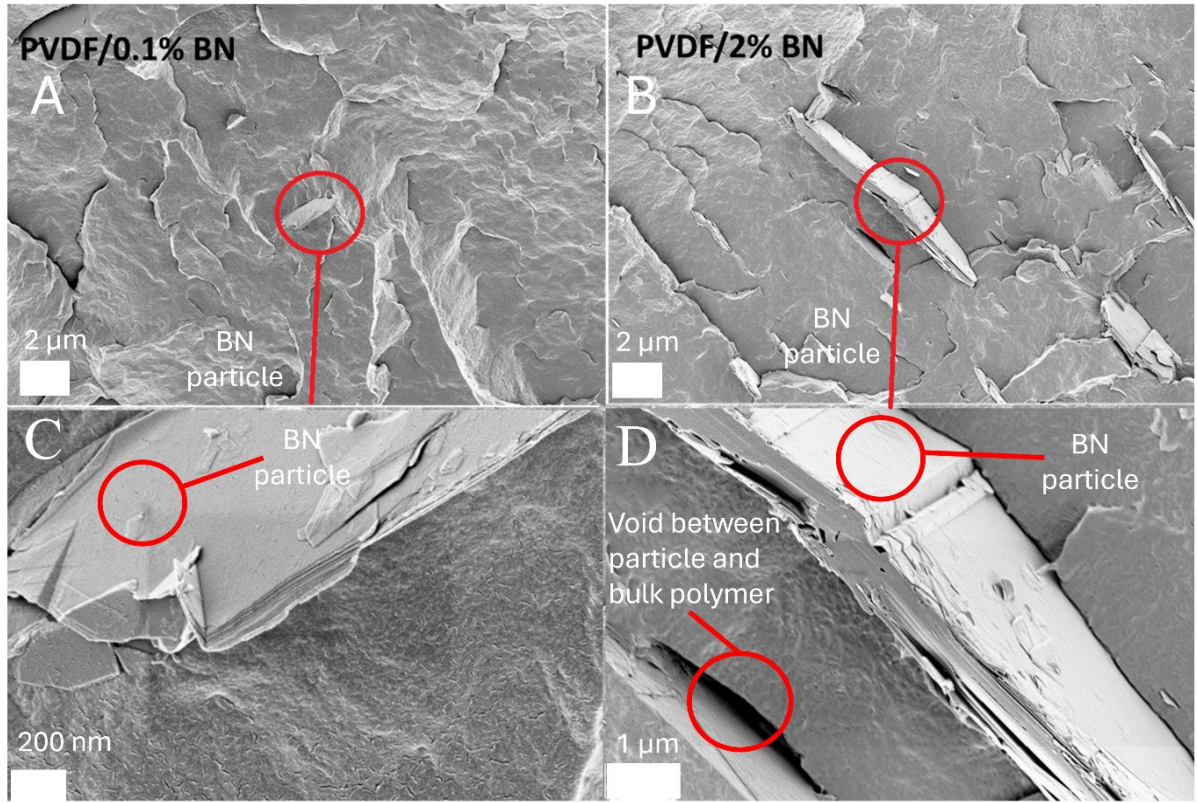

Figure 1 – SEM micrographs of freeze fractured (A) 0.1 wt. % BN and (B) 2 wt. % BN loadings in PVDF. BN particles are highlighted in red circles. Higher magnification images of the images in (A) and (B) are shown in (C) and (D) respectively. Voiding between the particle and the matrix can be observed, highlighted in D.

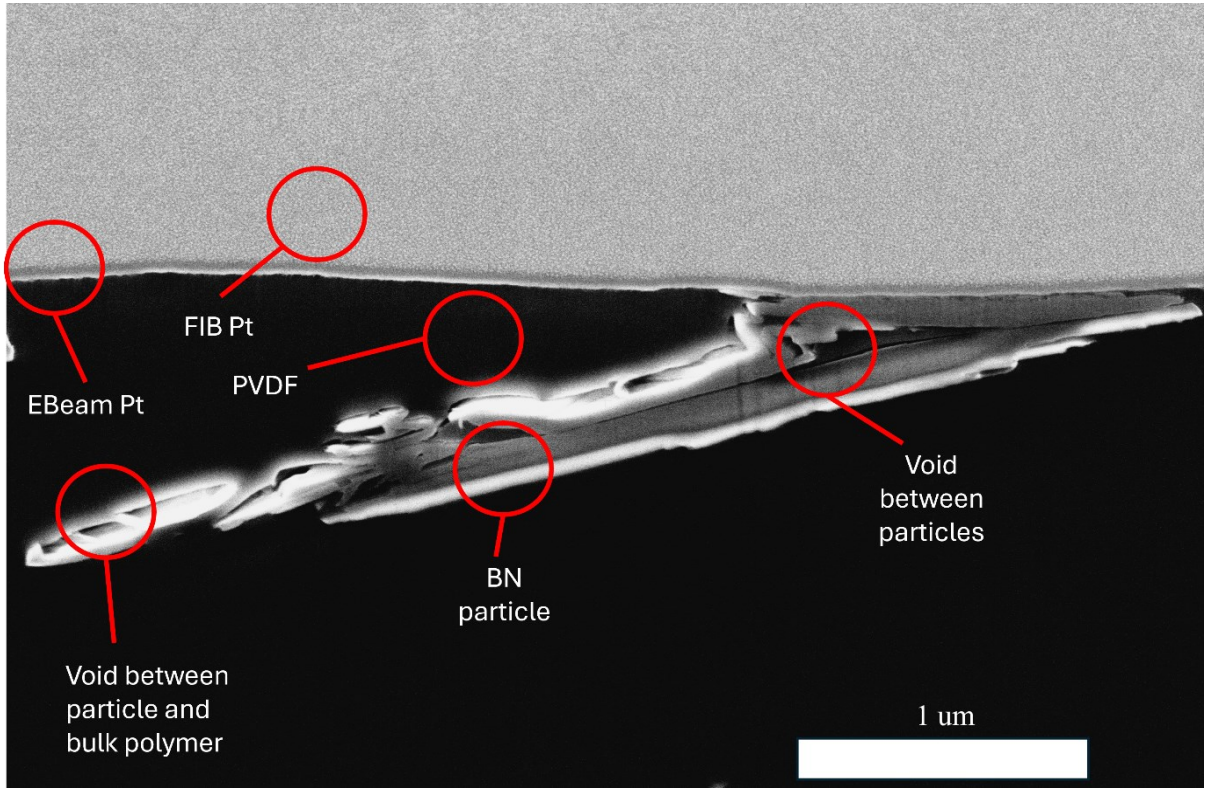

Figure 2 - SEM micrograph of Ga FIB cross section of BN particle (bright contrast) in PVDF bulk (dark contrast) from 5 wt. % BN loading with layer of FIB deposited Pt and electron beam deposited Pt protective layer above. Voiding between the particle and the matrix and voiding inside the particle is highlighted.

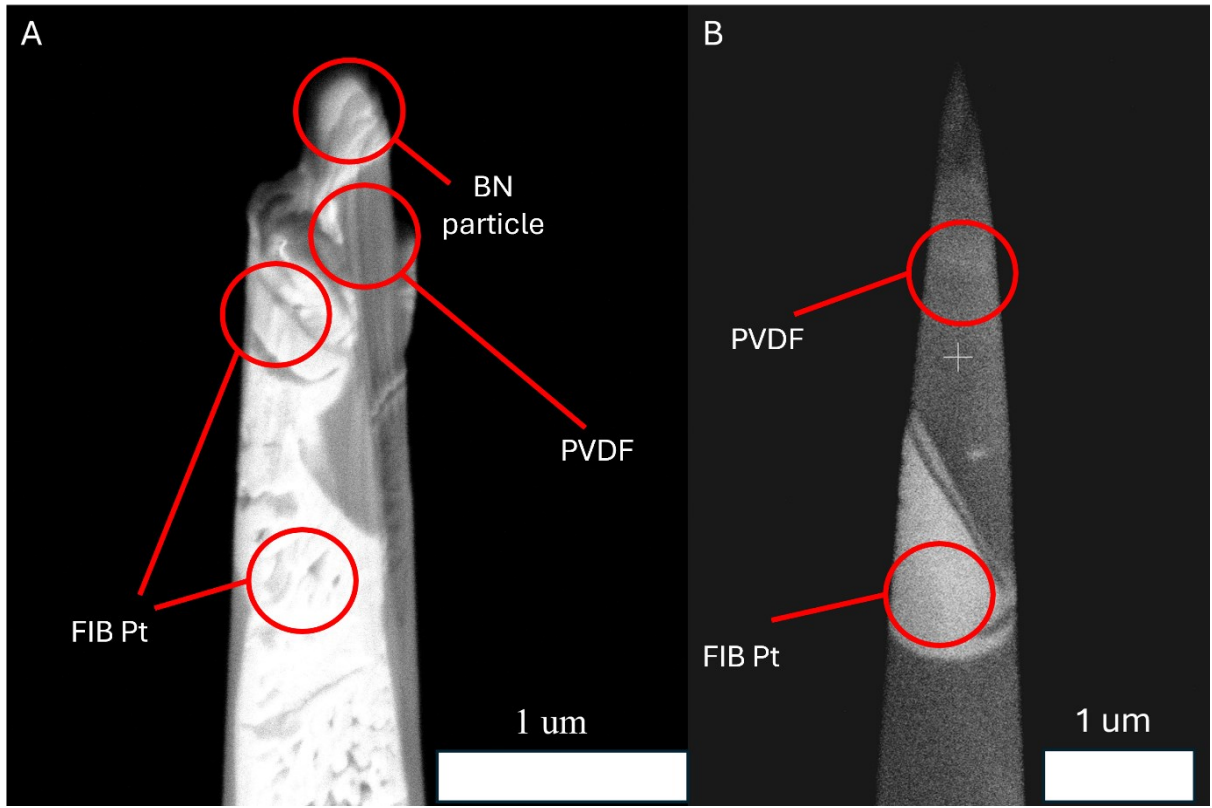

Figure 3 - SEM micrograph of Xe FIB sharpened APT specimens of (A) high contrast boron nitride particle at the apex and (B) low contrast bulk polymer. Scale bars 1 um.

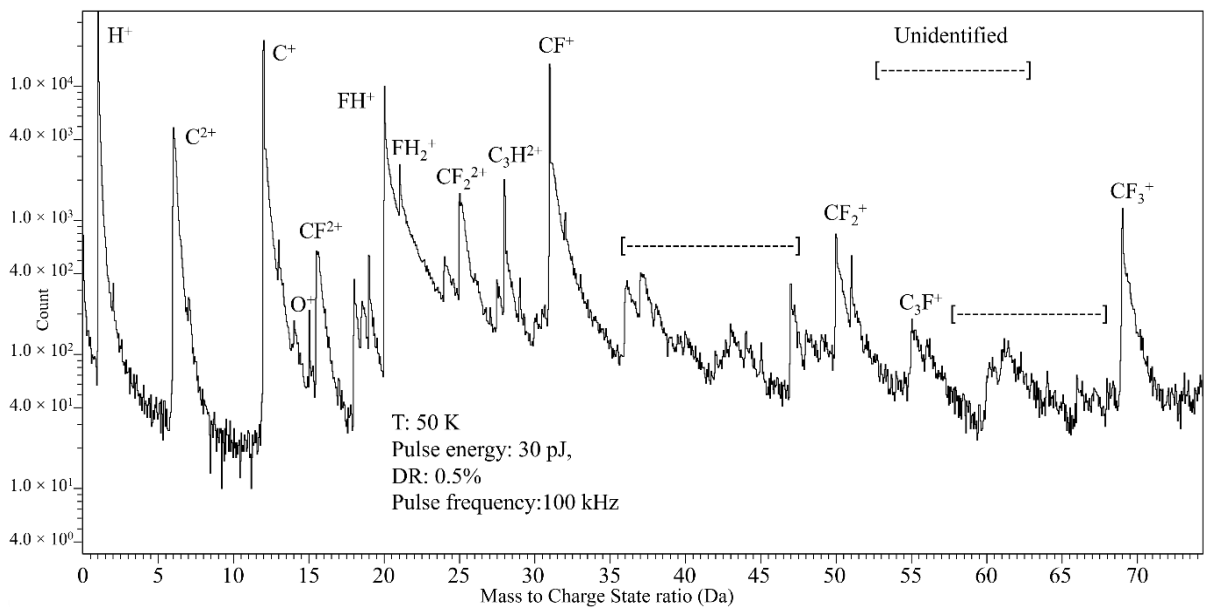

Figure 4 - Mass spectrum of room temperature Xe sharpened bulk polymer shown in Figure 3A.

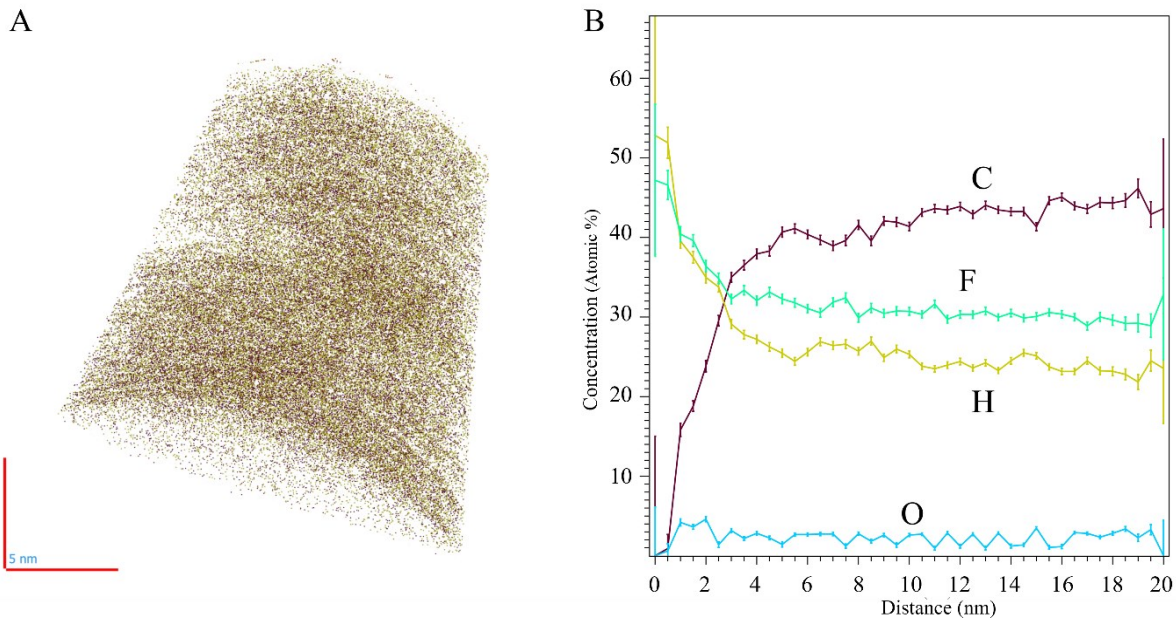

Figure 5 – (A) Atom map of room temperature Xe sharpened particle containing sample shown in Figure 3B showing homogeneous distribution of C species (brown), F species (green), H species (yellow) and O species (blue). Scale bar 5 nm. Not all ions shown for visualization purposes. (B) 1D depth concentration profile of decomposed elemental composition of Figure 5A with compound ions decomposed to elements. Nominal monomer ratio C:F:H 1:1:1.

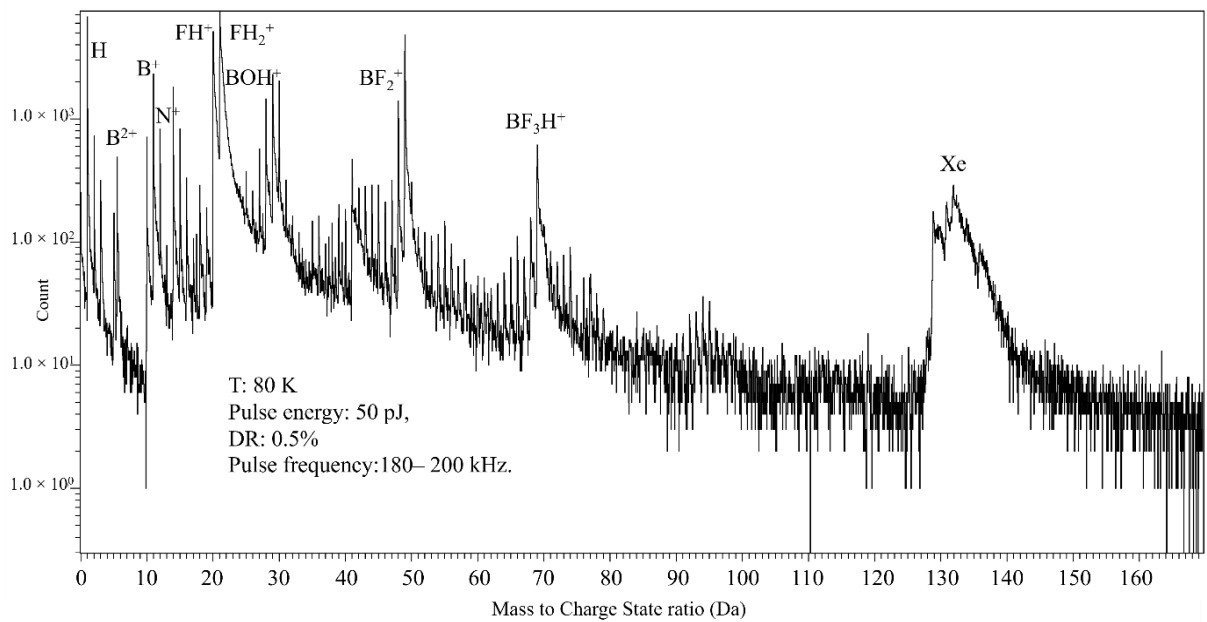

Figure 6 – Mass spectrum from room temperature Xe sharpened particle containing sample (0 Da to 160 Da) shown in Figure 3A.

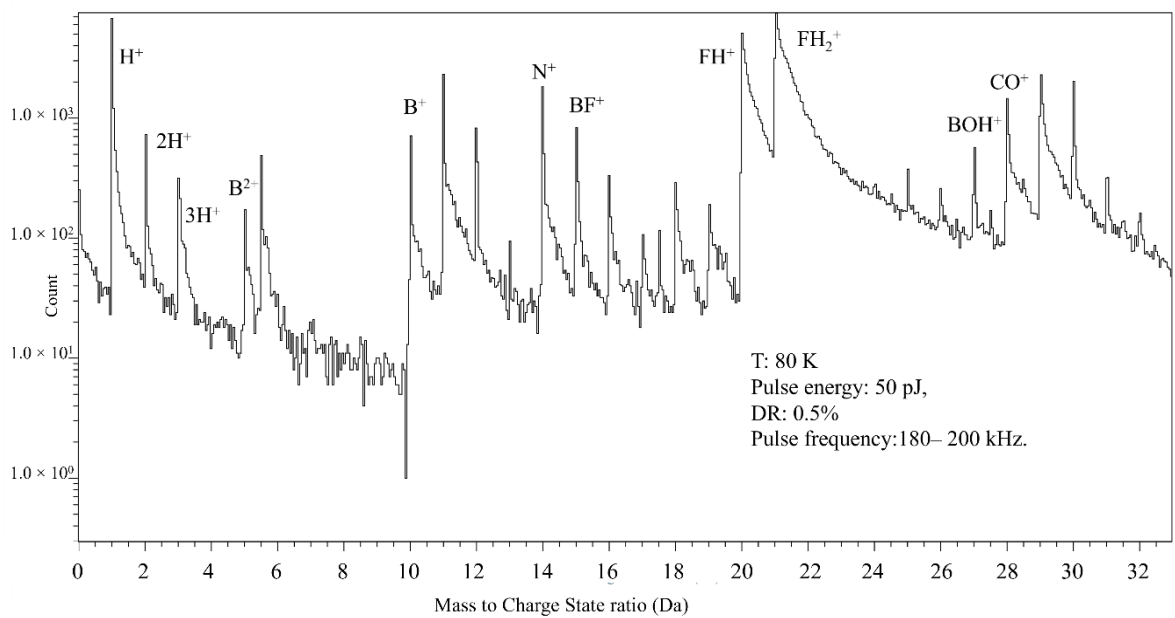

Figure 7 – Mass spectrum from room temperature Xe sharpened particle containing sample (0 Da to 32 Da) shown in Figure 3A.

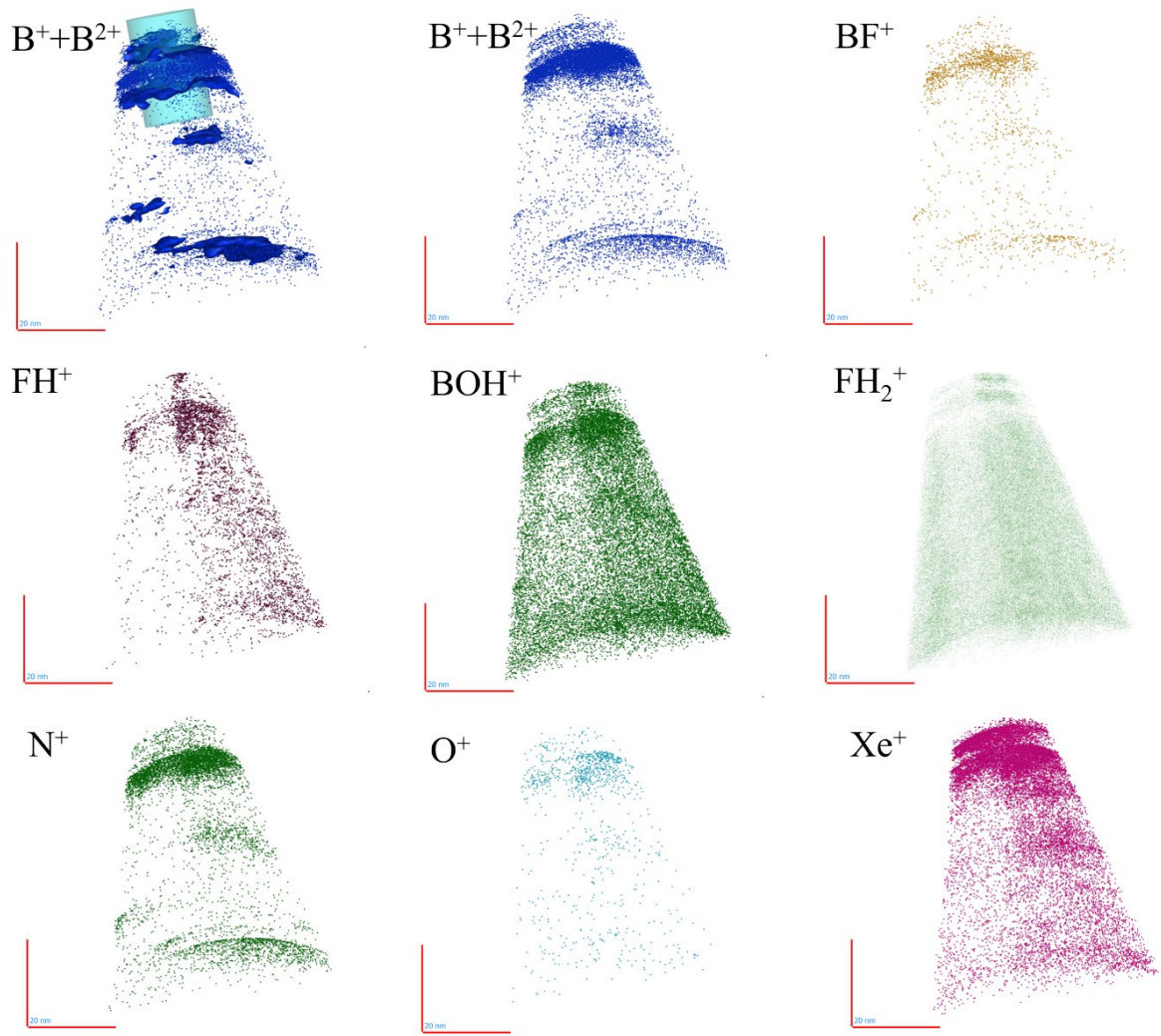

Figure 8 – Atom maps for room temperature Xe sharpened particle containing sample in Figure 3A. 2 at.% B isoconcentration surface used in top left atom map.

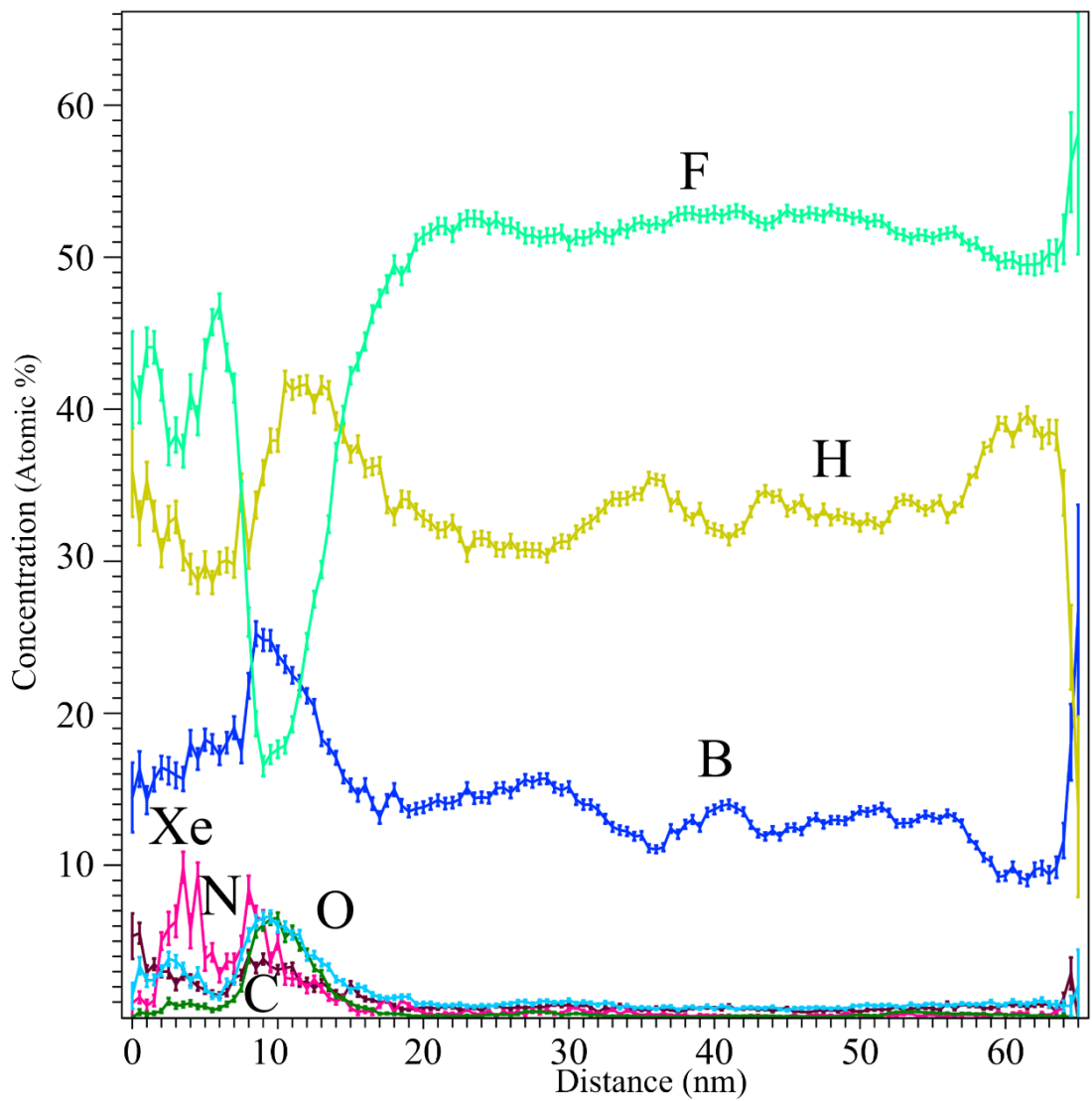

Figure 9 – A 1D concentration profile decomposed elemental composition in the analysis (Z) direction for room temperature Xe sharpened particle containing sample in Figure 8. Profile generated from the full length of the sample. F species (green), H species (yellow), boron species including all potential boron peaks (dark blue), Xe species (pink), N species (green), C species (brown), O species (light blue).

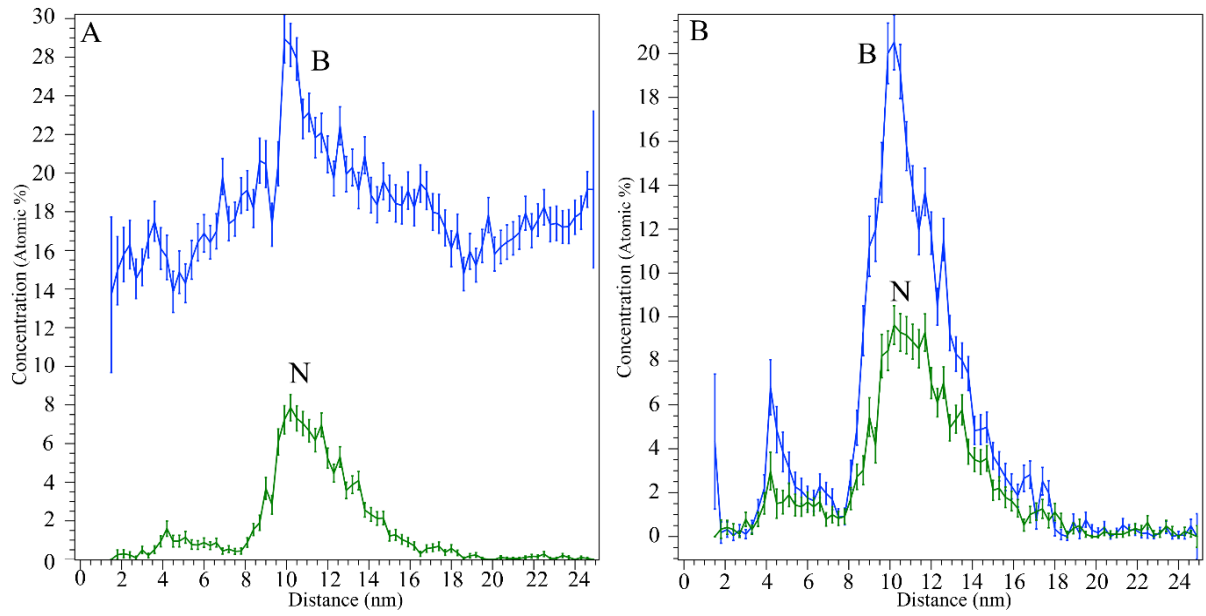

Figure 10 – (A) Decomposed 1 D concentration profile of consolidated potential B species and $N^+$ species ($B^+$, $B^{2+}$, $BF^+$, $N^+$, $N_2^+$, $BF_2^+$) across Region of Interest in Figure 8. (B) 1D concentration profile of unambiguous B species and N species ($B^+$, $B^{2+}$, $BF^+$, $N^+$) across Region of Interest in Figure 6. B species (dark blue), N species (green).

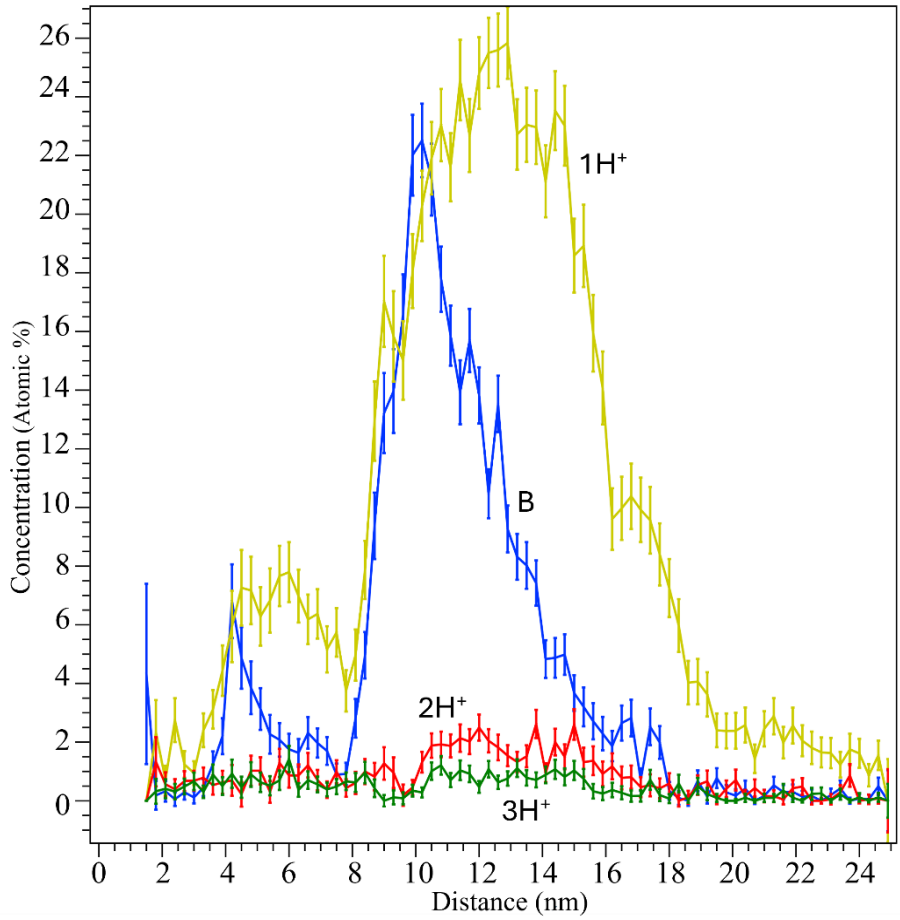

Figure 11 - 1D concentration profile of B ($B^+$ and $B^{2+}$) and $1H^+$, $2H^+$ and $3H^+$ species across cylindrical Region of Interest in Figure 6. B species (dark blue), $1H^+$ (yellow), $2H^+$ (red) and $3H^+$ (green).

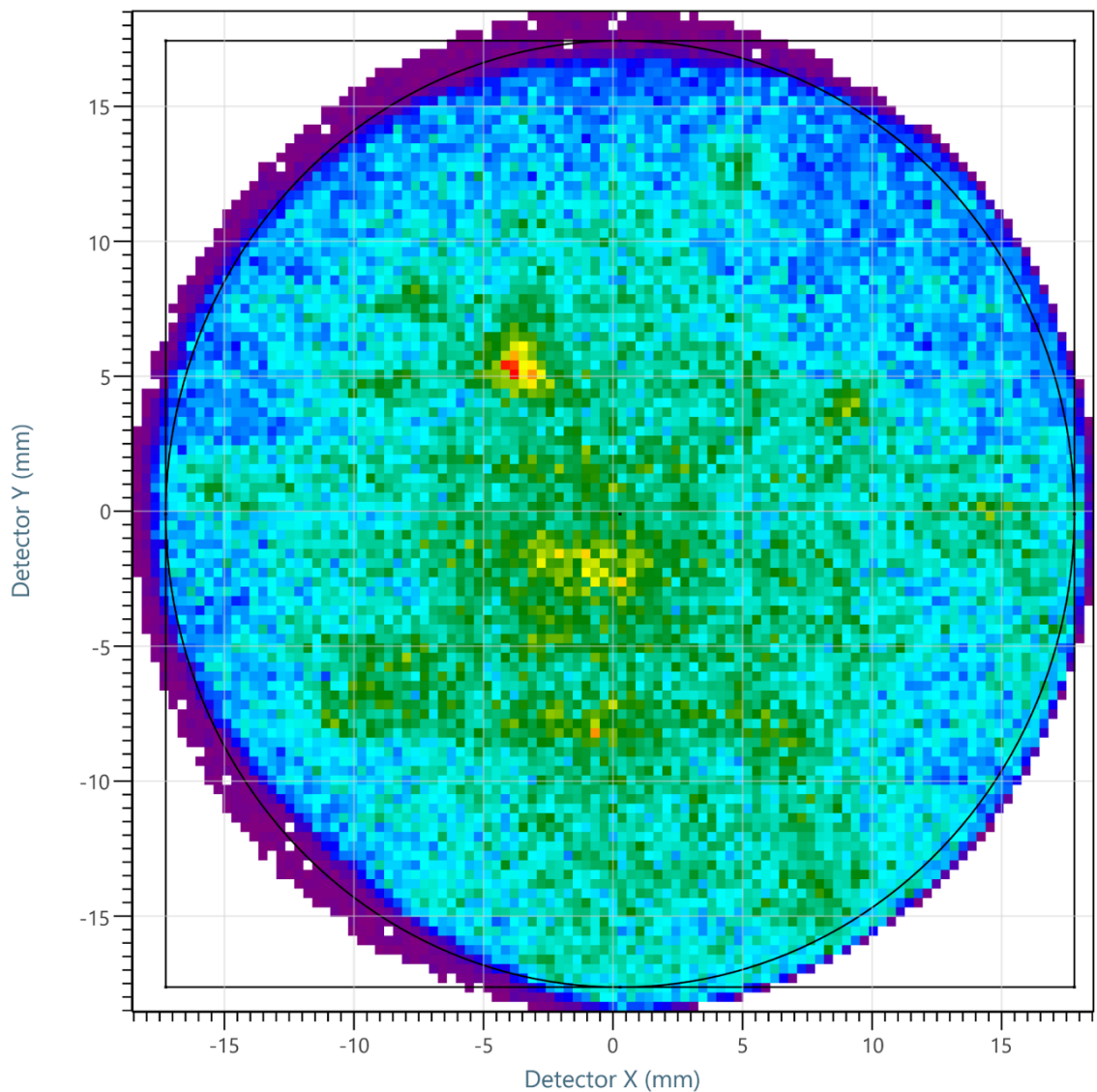

Figure S1 – Detector map from bulk polymer sharpened by room temperature Xe. The detector map of the bulk polymer sample sharpened by room temperature Xe FIB has some regions of high intensity but is reasonably uniform as would be expected of a bulk material.

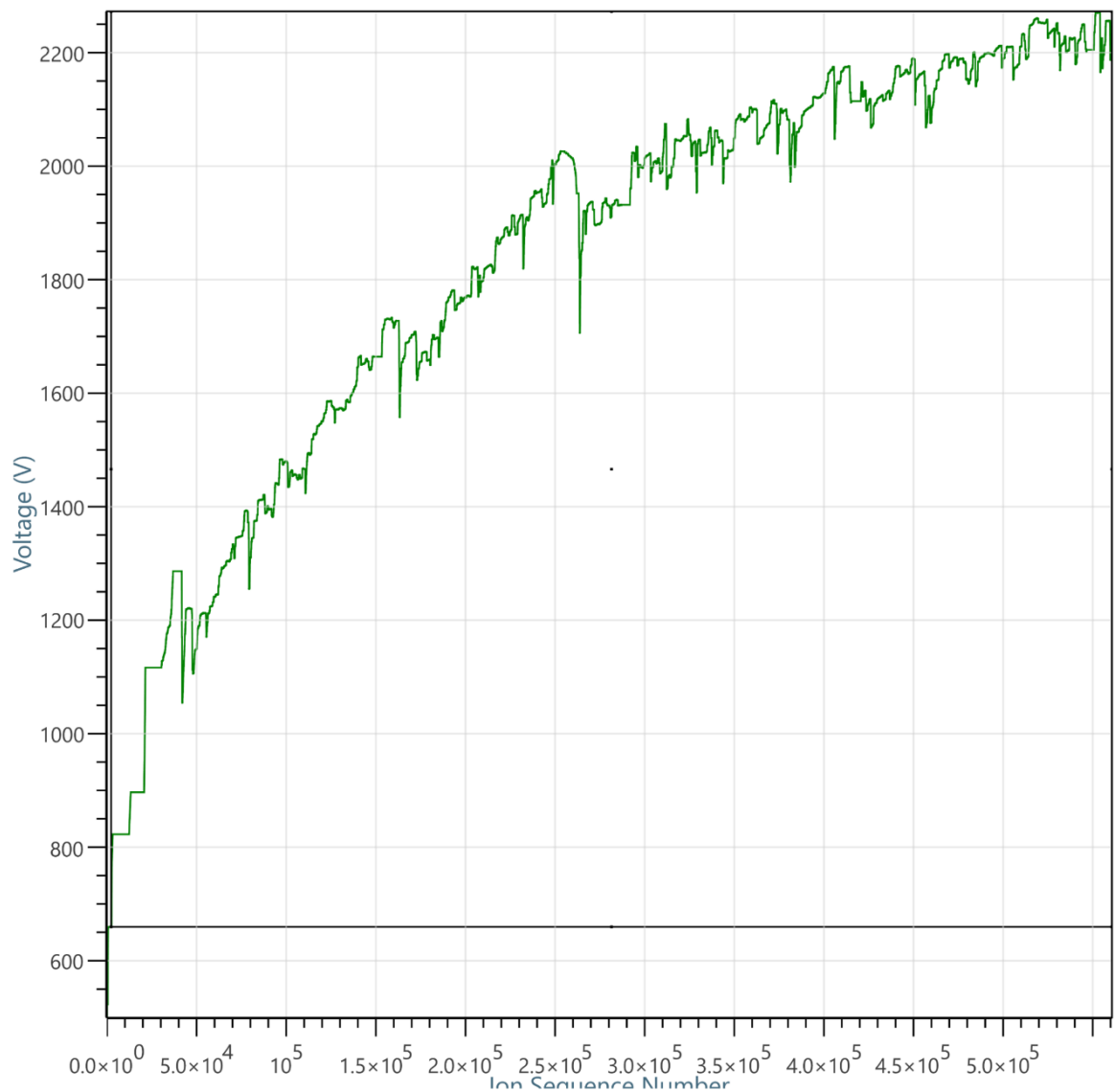

Figure S2 – Voltage curve from bulk polymer sharpened by room temperature Xe. Voltage curve follows general trend of increasing voltage with increasing ion count as would be expected of an increasing shank angle.

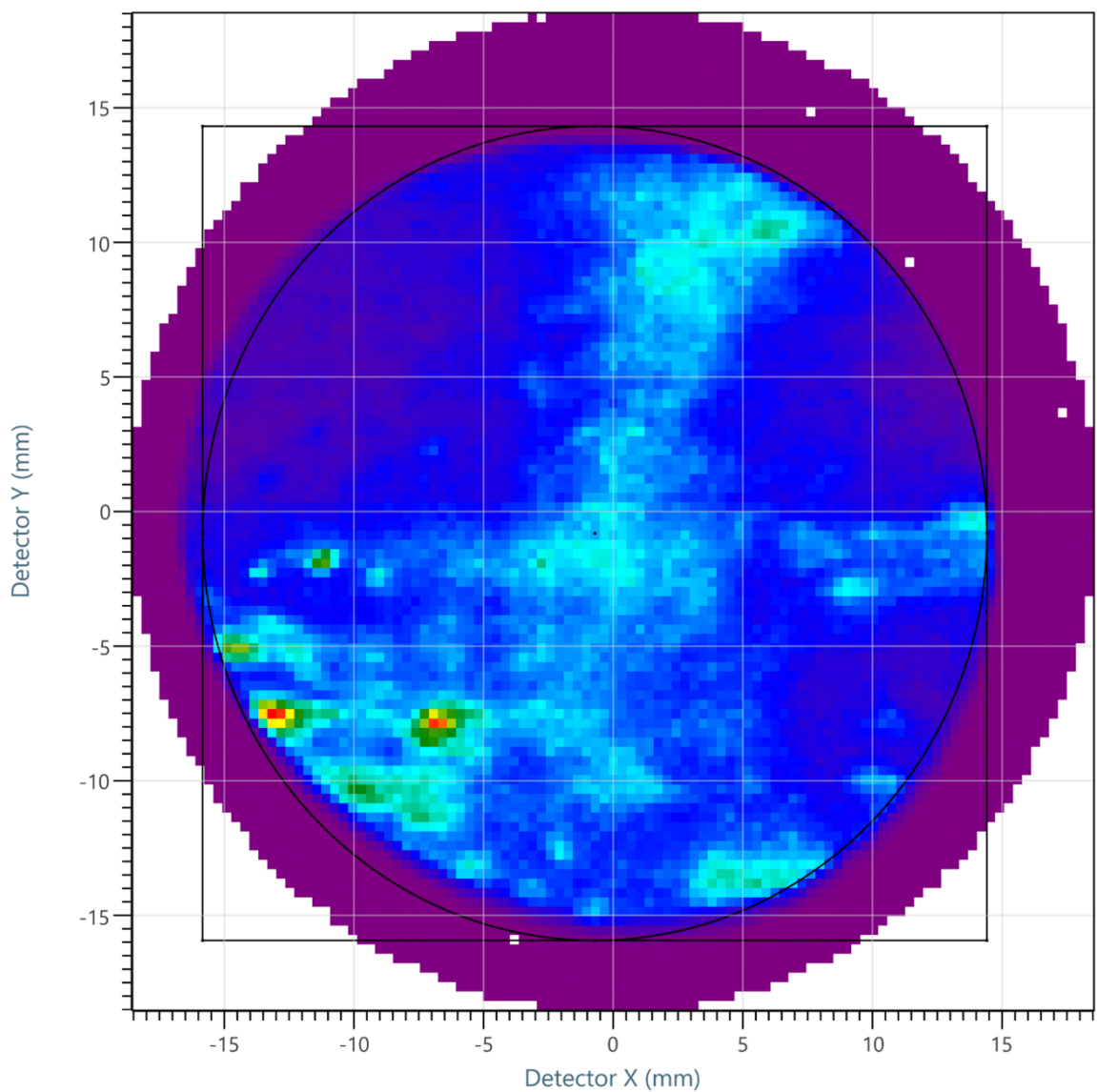

Figure S3 – Detector map from sample containing BN particle sharpened by room temperature Xe. The detector map of the sample containing the BN particle prepared by Xe FIB is more inhomogenous than that of the bulk sample, indicating less uniform behaviour in evaporation from the particle regions compared to the bulk.

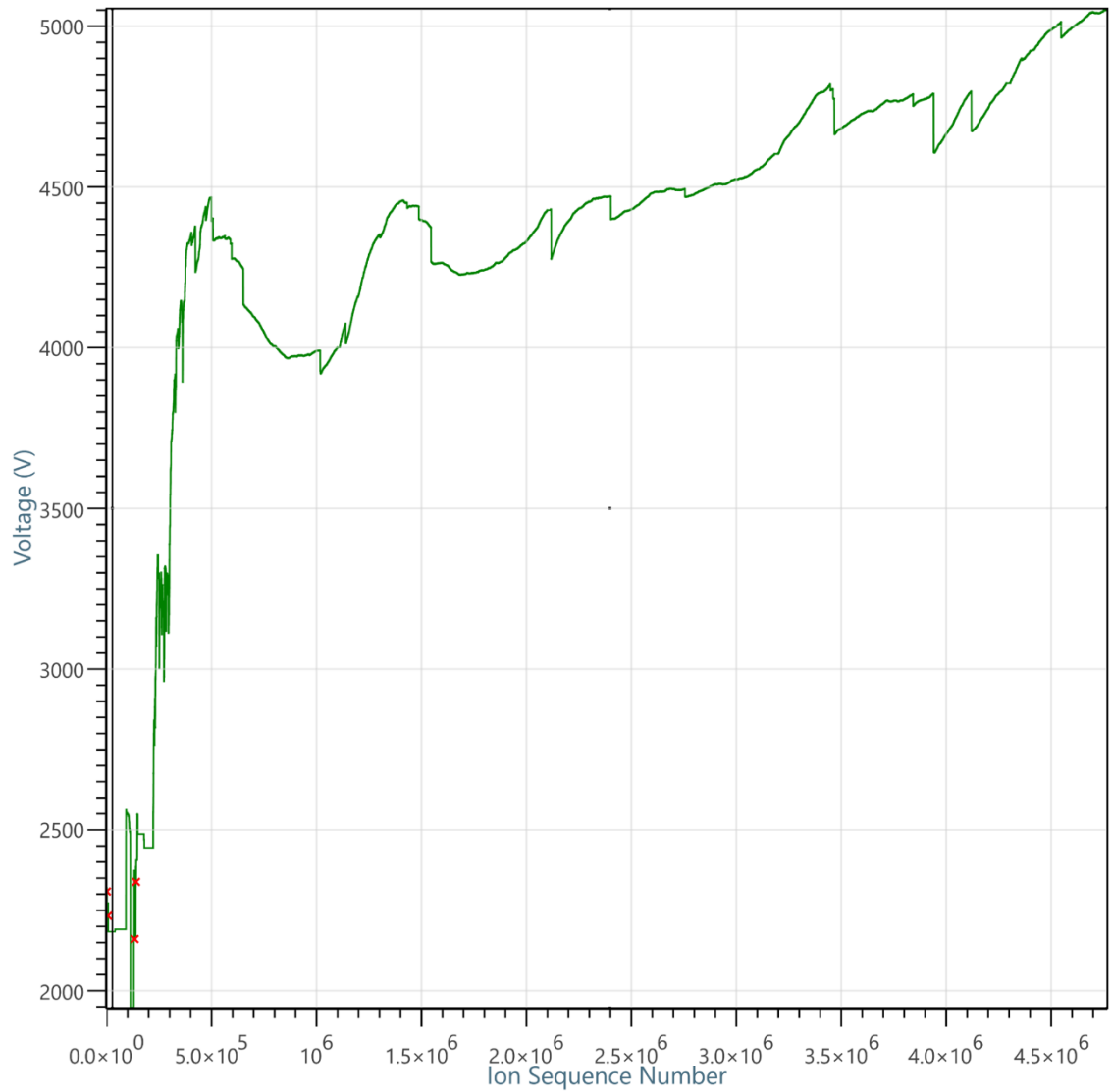

Figure S4 – Voltage curve from sample containing BN particle sharpened by room temperature Xe. The voltage curve has a general trend of increasing voltage with increasing ion count but also has jumps in voltage during analysis. This indicates either changes in required voltage due to changs in composition of the region being evaporated or micro-fracture events or potentially a combination of the two. The initial voltage required is significantly higher than that of the bulk polymer sample and this can be attributed to both the differences in composition and also the larger initial diameter of the sample

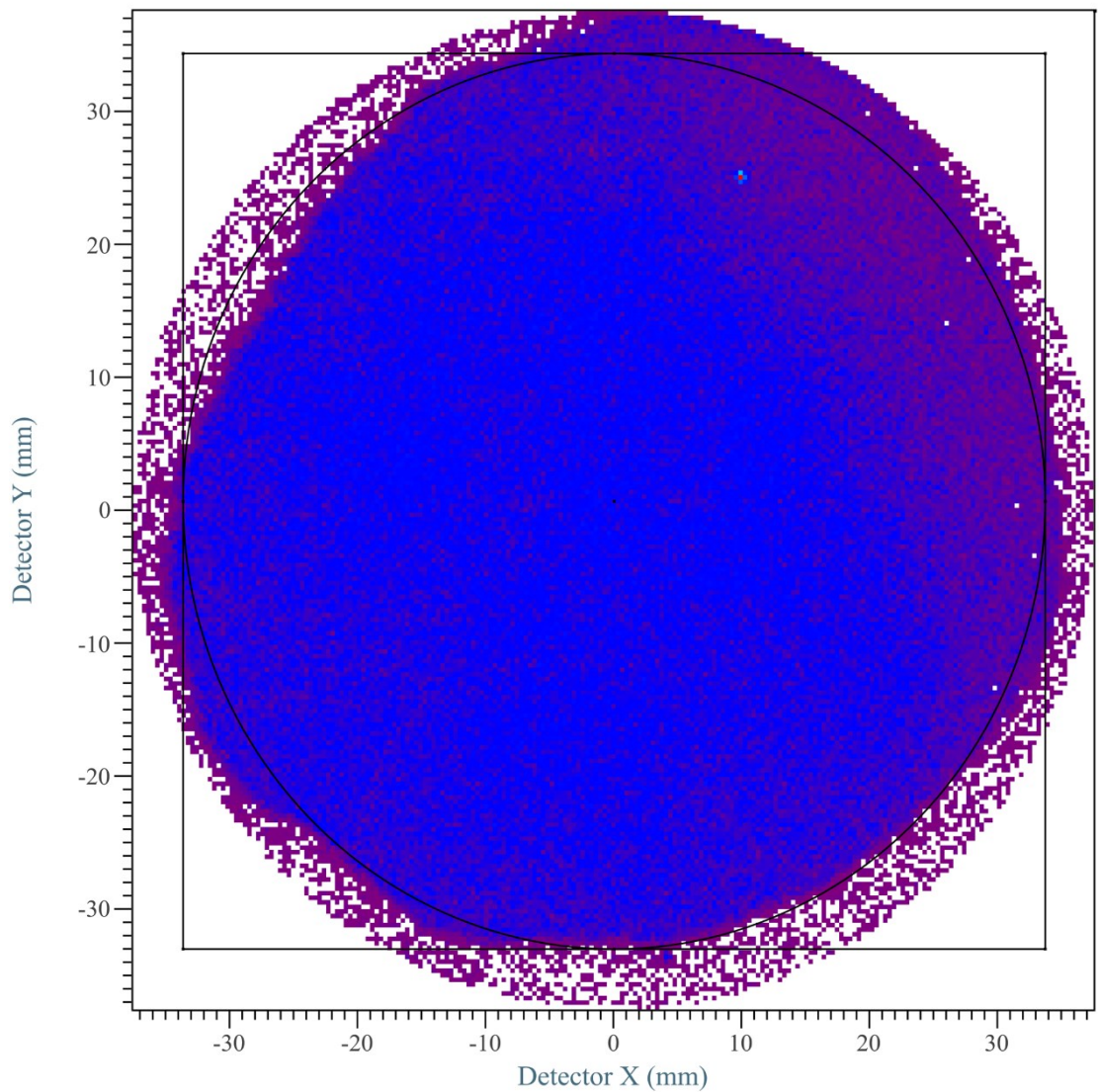

Figure S5 – Detector map of bulk polymer sample sharpened by room temperature Ga and charged with $D_2$. The detector map of the bulk polymer sample sharpened by room temperature Ga FIB and $D_2$ charged appears to be homogeneous, more so than that of the sample sharpened by room temperature Xe FIB shown in Figure S1.

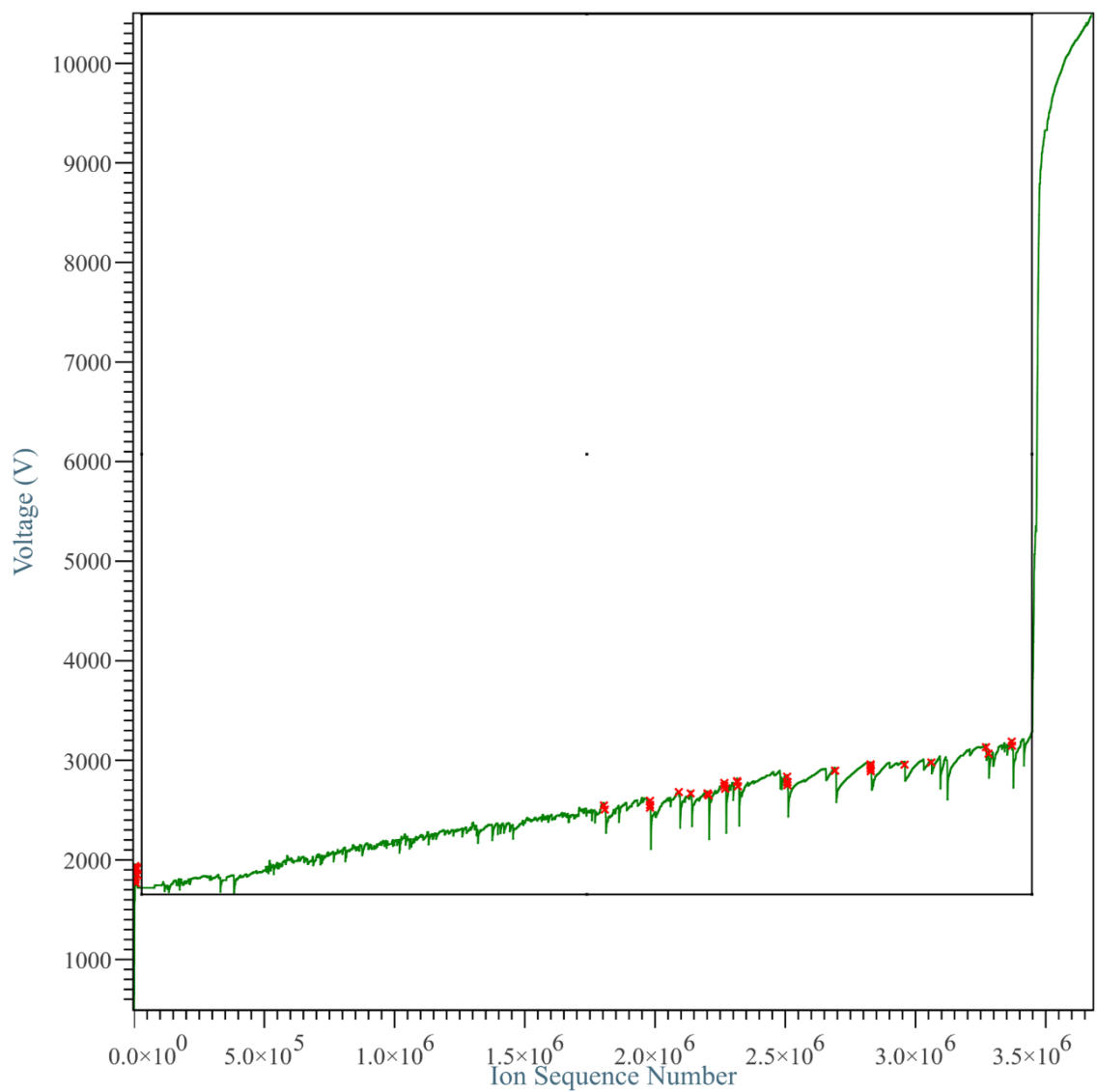

Figure S6 – Voltage curve from bulk sample sharpened by room temperature Ga sharpened and charged with $D_2$. The voltage curve has a general trend of increasing voltage with increasing ion count but also has jumps in voltage during analysis, similar to that of the Xe room temperature sample show in Figure S4. The voltage curve also spans a similar voltage range as Figure S4 and shows a sample fracture and associated voltage rise at the end of the analysis.

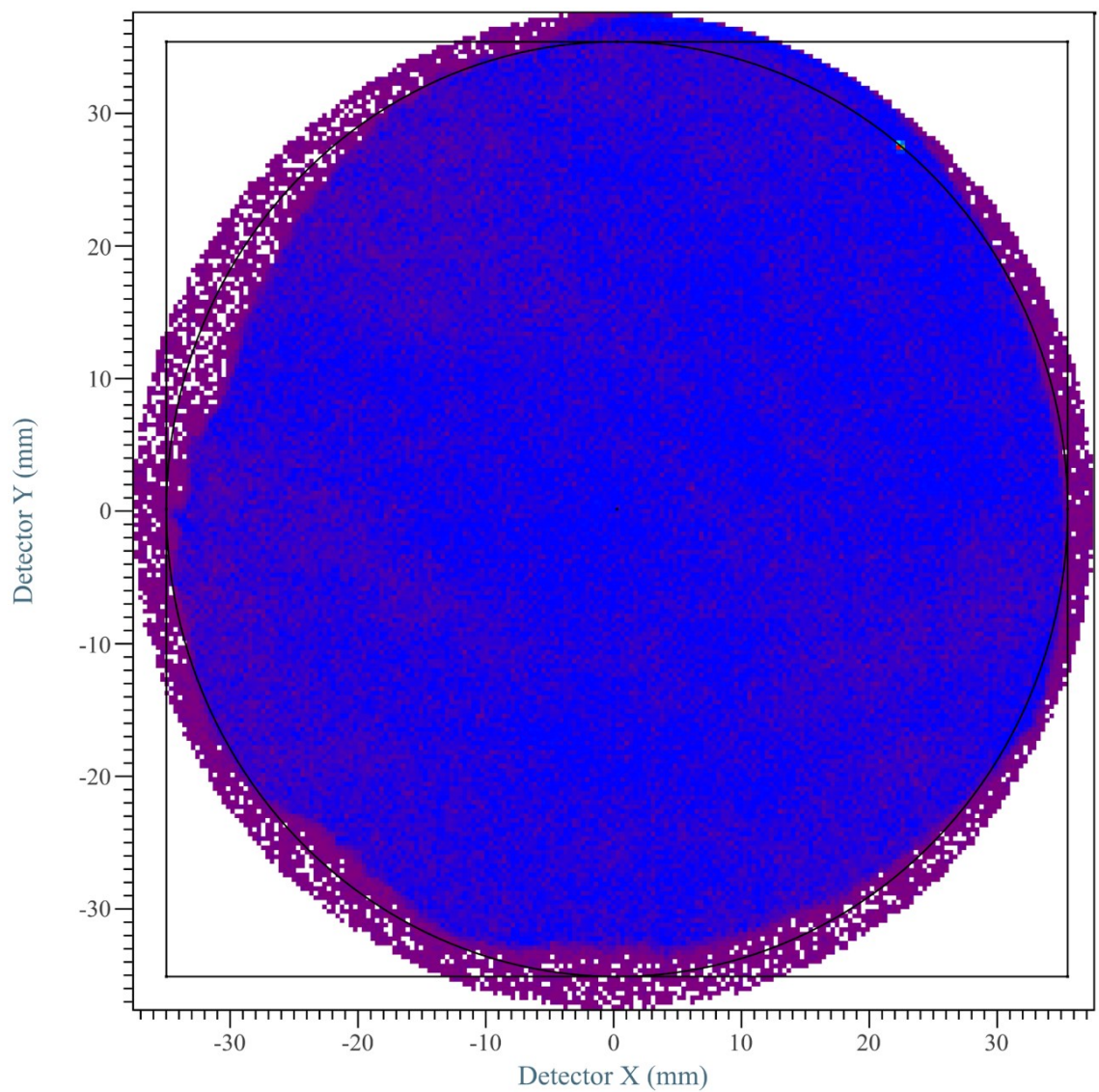

Figure S7 – Detector map of bulk polymer sharpened by cryogenic Ga and D₂ charged. The detector map of the bulk polymer sample sharpened by cryogenic Ga FIB and D₂ charged appears to be homogeneous, very similar to that of the room temperature Ga FIB shown in Figure S5. This could imply that the effect of Ga sharpening on the sample may not be changed significantly due to temperature changes when this information is combined with the lack of change in mass spectra between samples prepared between the two workflows.

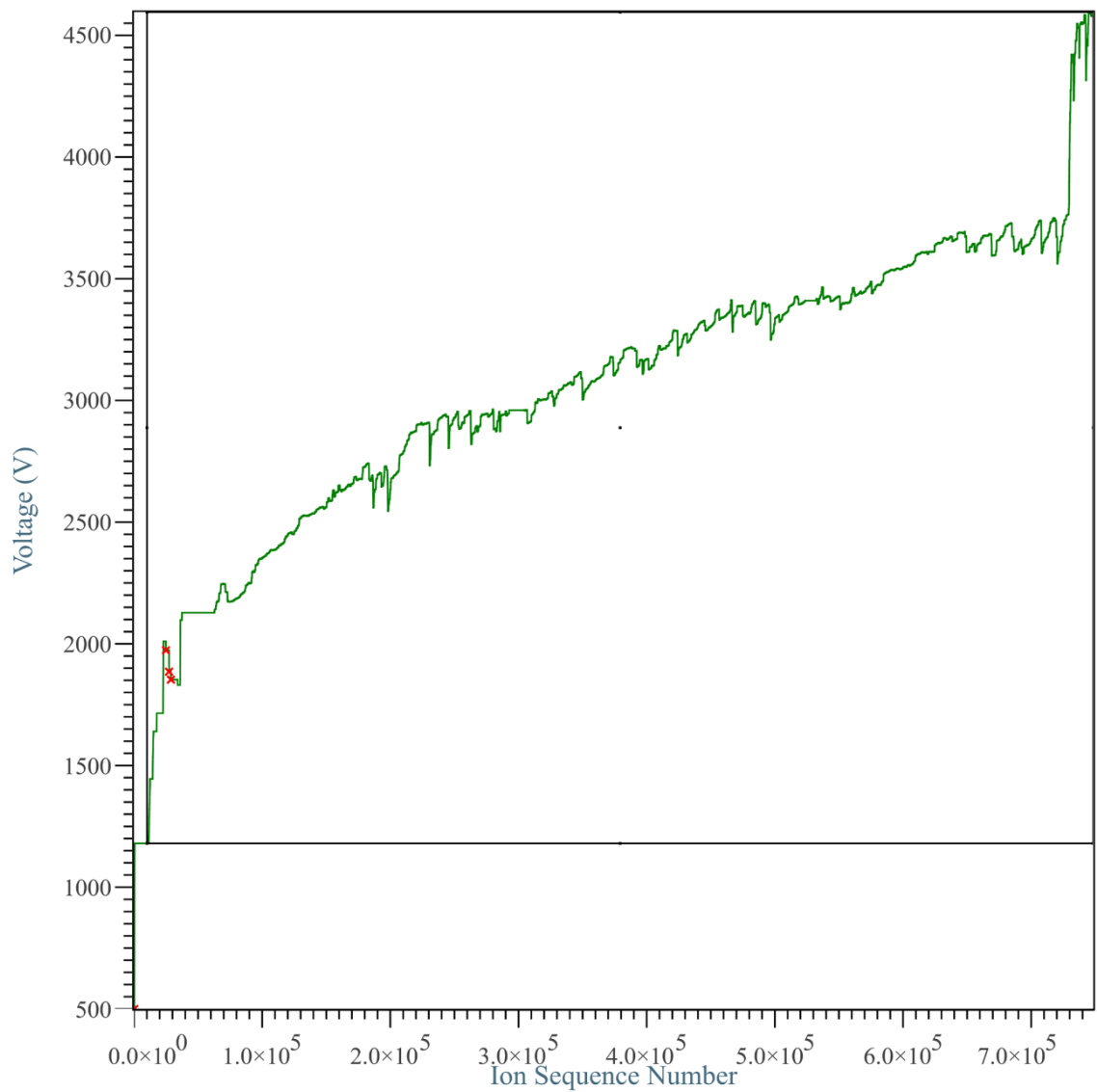

Figure S8 - Voltage curve from bulk polymer sharpened by cryogenic Ga FIB sharpened and D$_2$ charged. The detector map of the bulk polymer sample sharpened by room temperature Ga FIB and D$_2$ charged appears to be homogeneous, more so than that of the sample sharpened by room temperature Xe FIB shown in Figure S1.

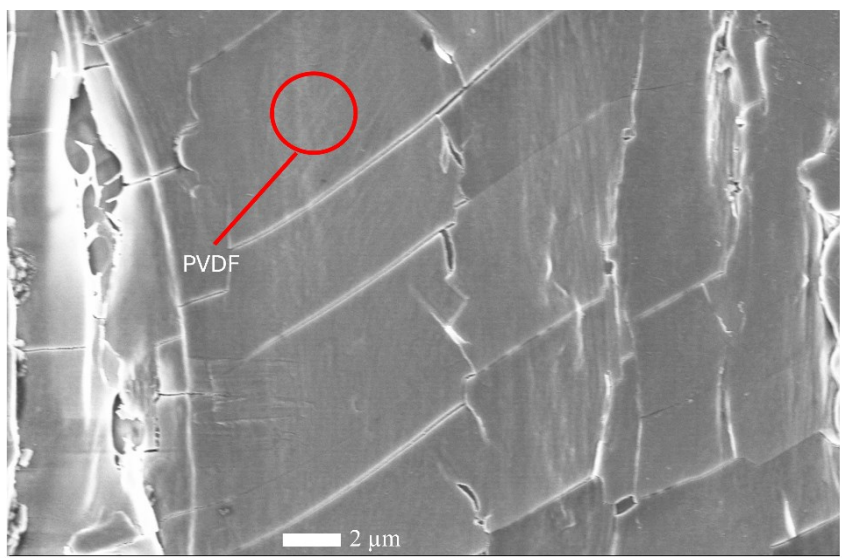

Figure S9 - SEM image of planar surface of 5% BN in PVDF used for EDS analysis. Red circle indicates location of EDS spot analysis.

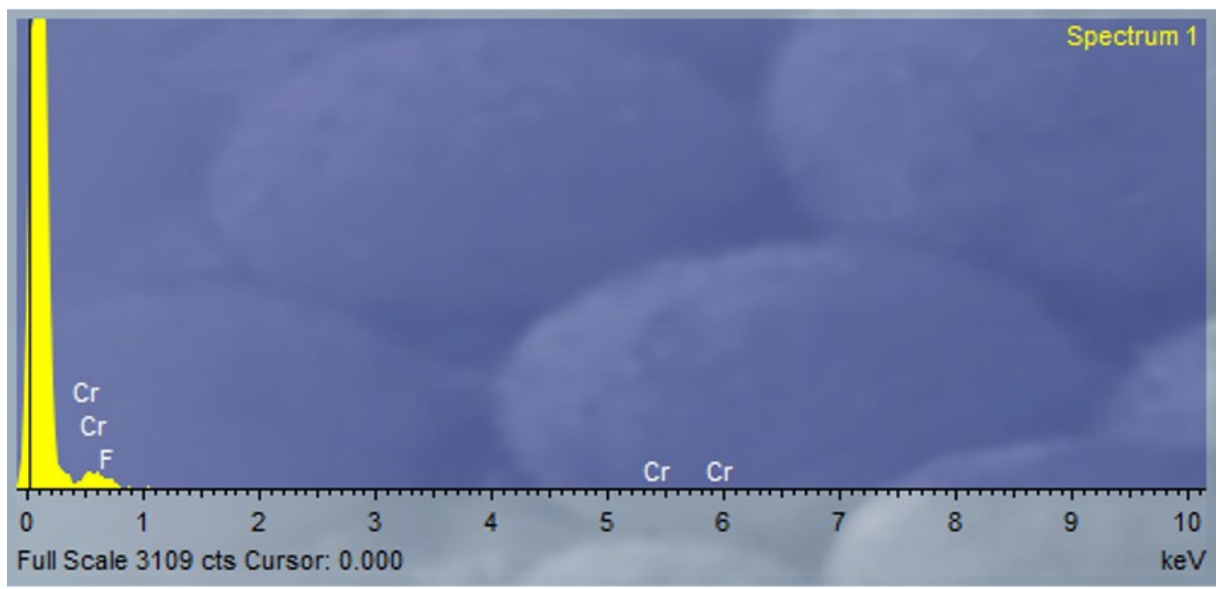

Figure S10 – EDS spectra of spot shown in Figure S9. No peaks of N or B observed. Small peak of F auto detected and likely linked with bulk polymer. Small amounts of Cr autodetected which are likely artefacts.

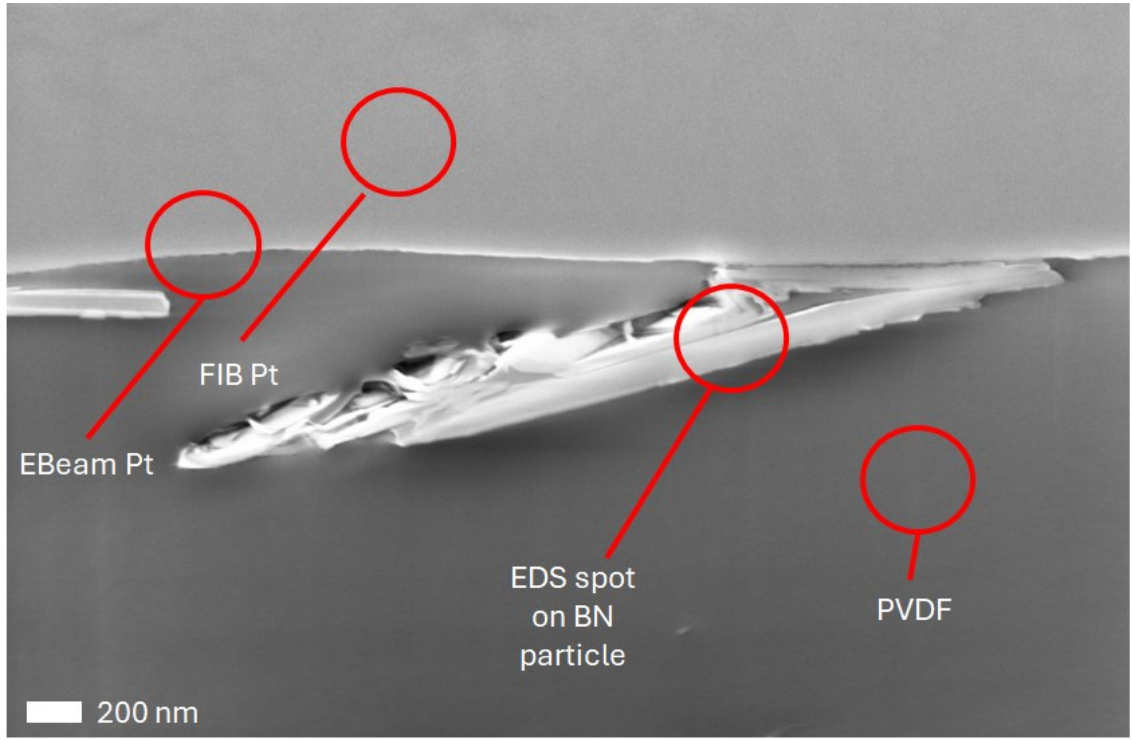

Figure S11 - SEM image of FIB cross section of 5% BN in PVDF used for EDS analysis. Red circles mark regions of interest.

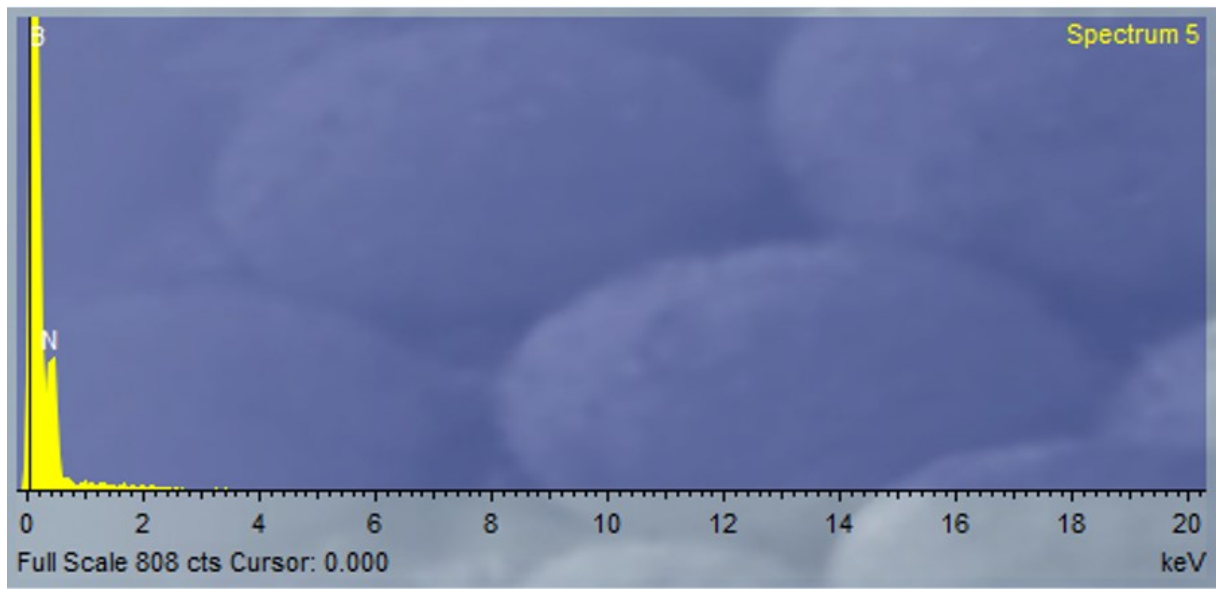

Figure S12 – EDS spectra of spot shown in Figure 11. Small peak of B auto detected, peak of N auto detected

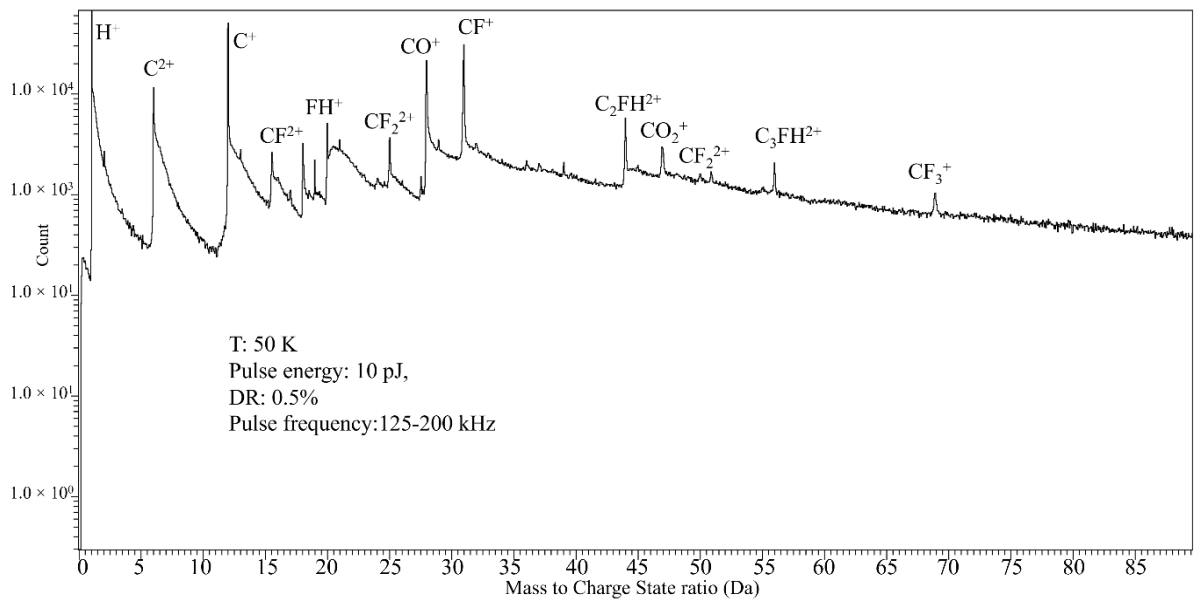

Figure S13 – Mass spectra of bulk polymer sharpened with room temperature and $D_2$ charged. The mass spectra of the room temperature Ga sharpened and D2 charged polymer sample shows multiple C, F and H containing compound ion which can be linked to various fragments of the PVDF polymer chain.

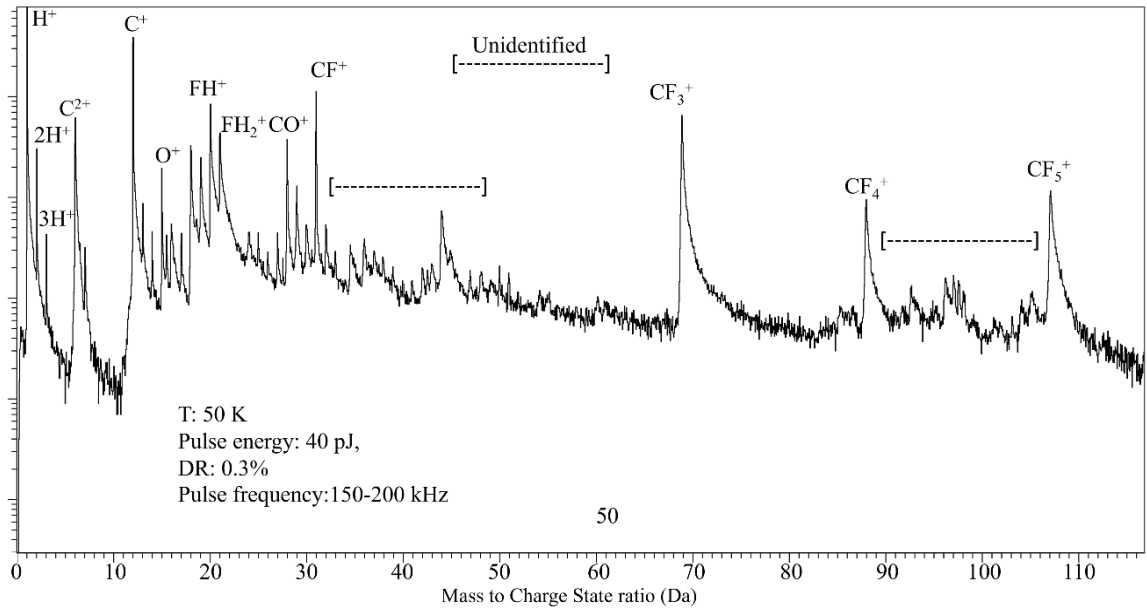

Figure S14 – Mass spectra of bulk polymer sharped with Ga FIB and D$_2$ charged. The mass spectra of the cryogenic Ga sharpened and D2 charged polymer sample shows multiple C, F and H containing compound ion which can be linked to various fragments of the PVDF polymer chain. The spectra is similar to that from the room temperature Ga sharpened sample, with some heavier CF based compound ions.

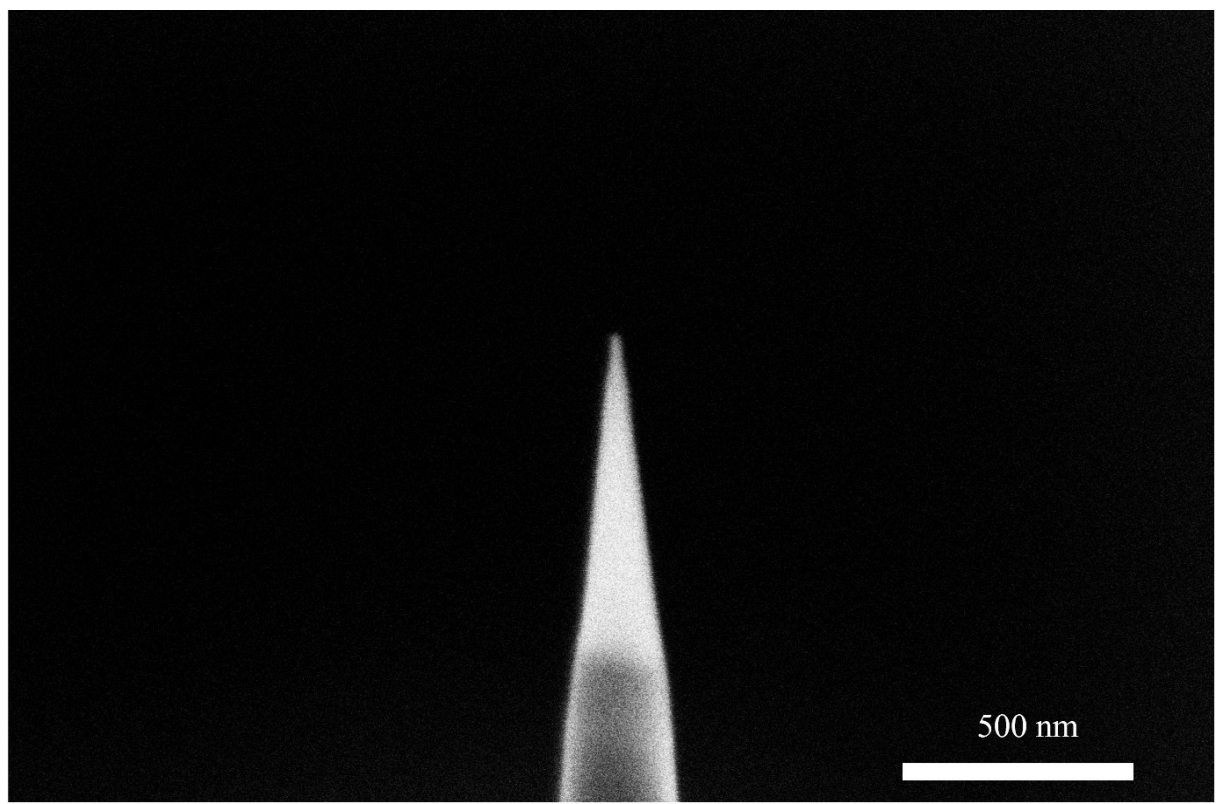

Figure S15 – SEM micrograph of bulk polymer sharpened with Ga FIB before $D_2$ charging. Scale bar 500 nm.

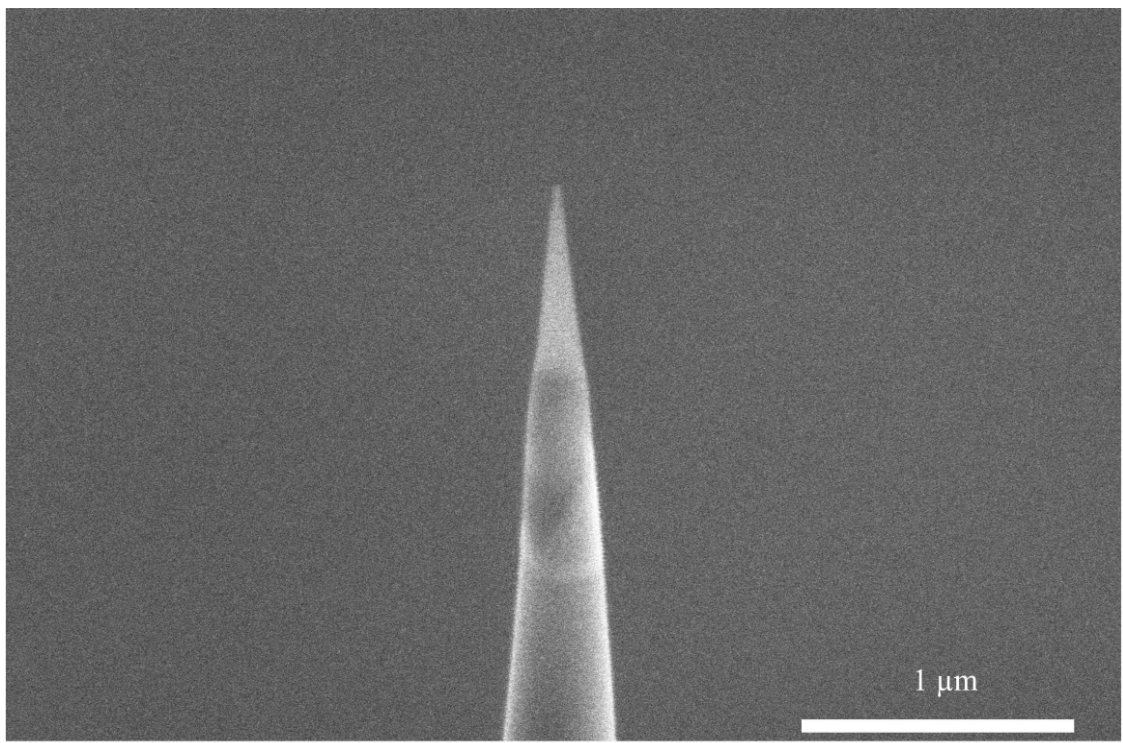

Figure S16 – SEM micrograph of bulk polymer sharpened with Ga FIB shown in Figure S15 after $D_2$ charging. No visible change in structure. Scale bar 1 μm.

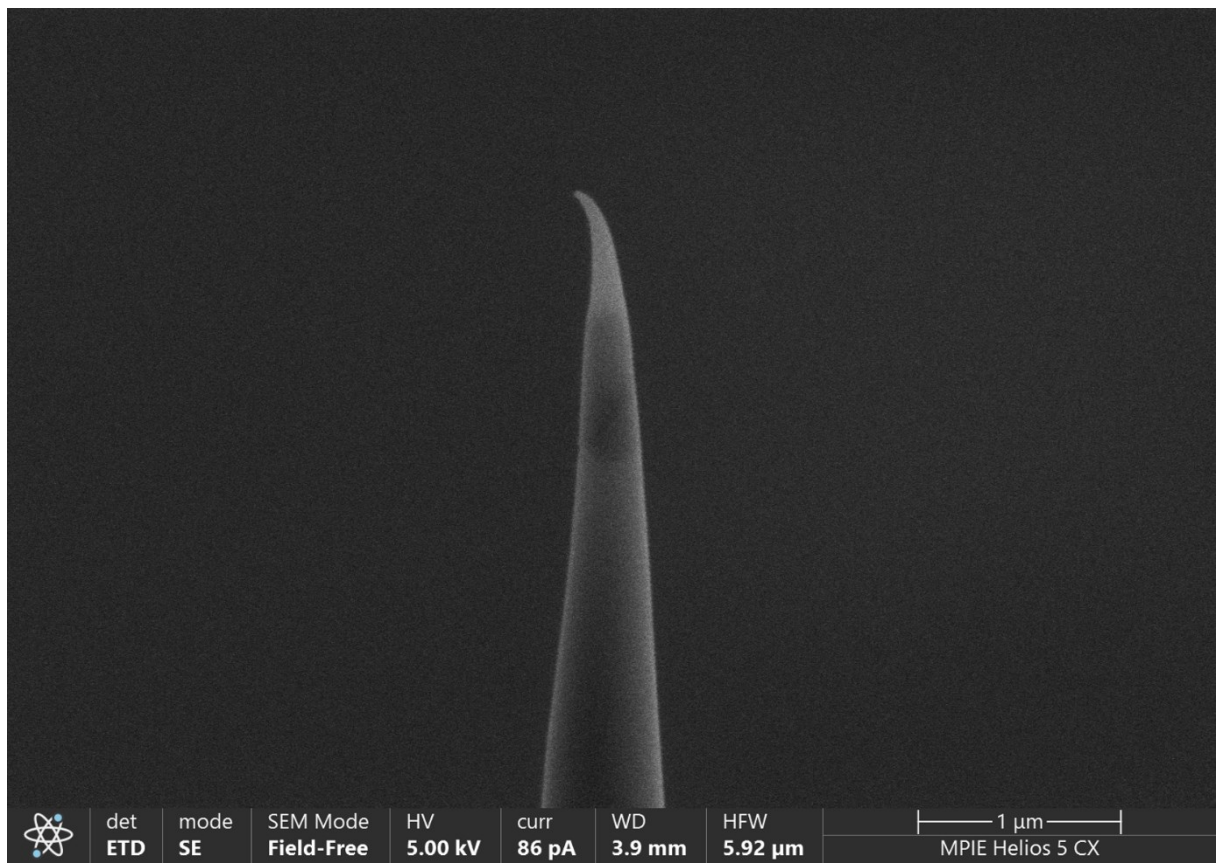

Figure S17 – SEM micrograph of bulk polymer sharpened with Ga after D2 charging then coated in-situ at cryogenic temperature with Cr. Scale bar 1 μm.

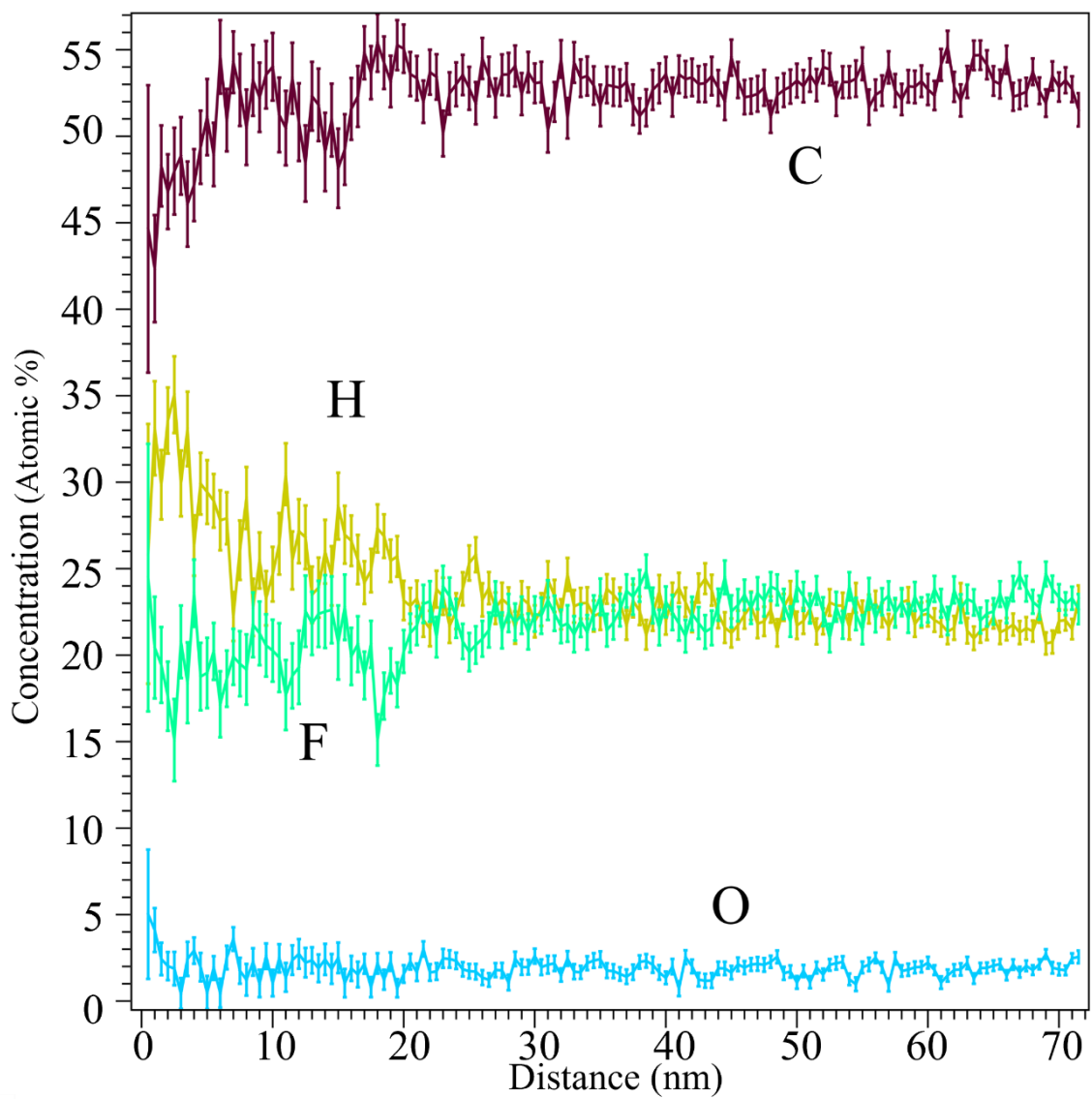

Figure S18 – 1D concentration profile of decomposed elemental composition of bulk polymer sharpened with room temperature Ga and then D2 charged. A 1D concentration profile in the analysis (Z) direction of decomposed C, F, H and O species shows an initial transition at the surface before the composition settles into a steady state plateau with C being the highest concentration species and H and F being approximately equal. O is at a consistent low level throughout.

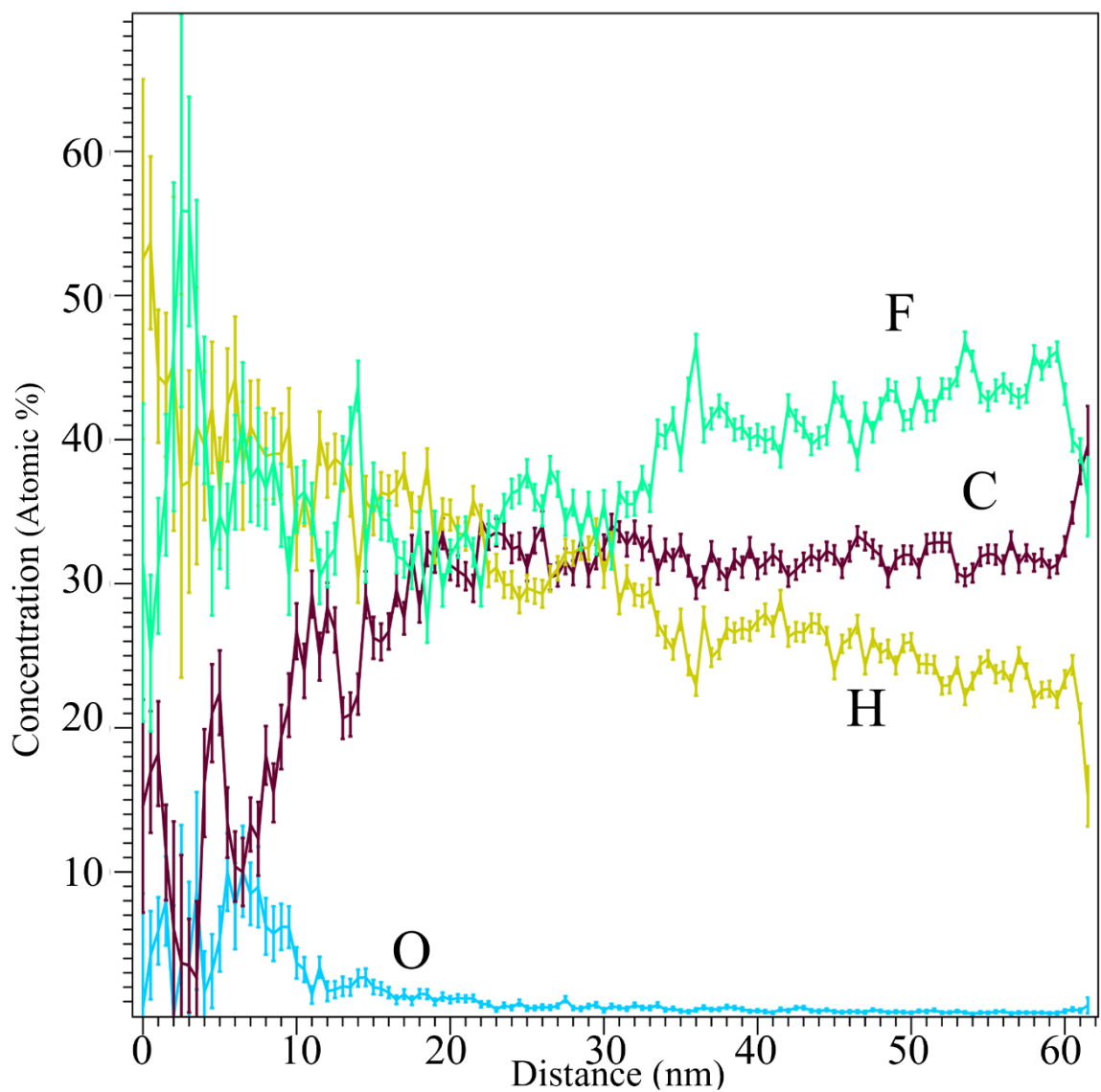

Figure S19 – 1D concentration profile of decomposed elemental composition of bulk polymer sharpened with cryogenic Ga and then $D_2$ charged. A 1D concentration profile in the analysis (Z) direction of decomposed C, F, H and O species shows an initial transition at the surface before the composition settles into a steady state plateau. F is the highest concentration, followed by C and then H. O peaks at the surface and then decreases with depth.

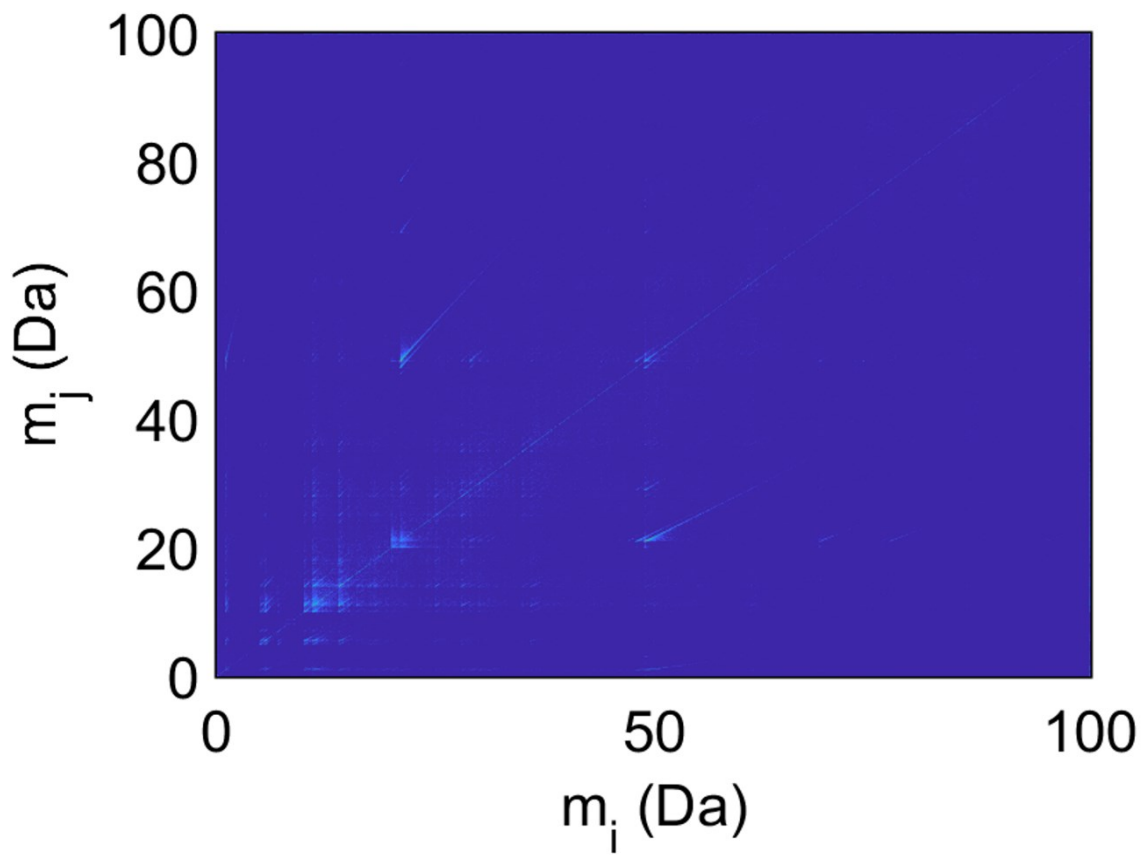

Figure S20 – Multiple hit correlation histogram of mass spectra up to 100 Da collected from bulk polymer sharpened with room temperature Xe. Histogram shows no visible dissociation tracks.

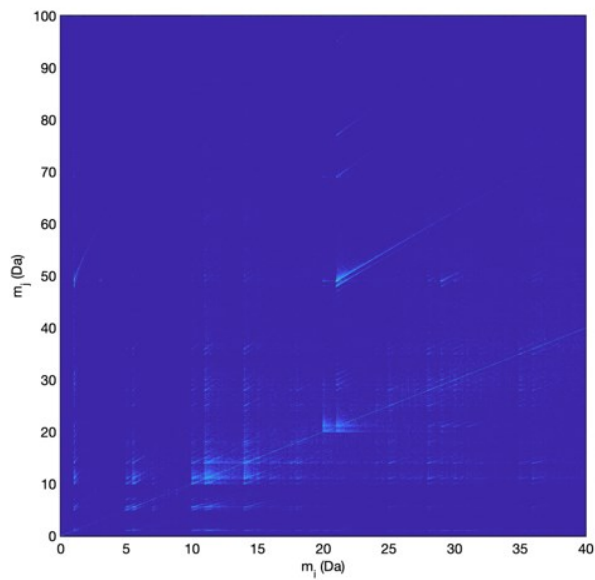

Figure S21 – Multiple hit correlation histogram of mass spectra from 0 Da to 40 Da collected from bulk polymer sharpened with room temperature Xe. Histogram shows no visible dissociation tracks.

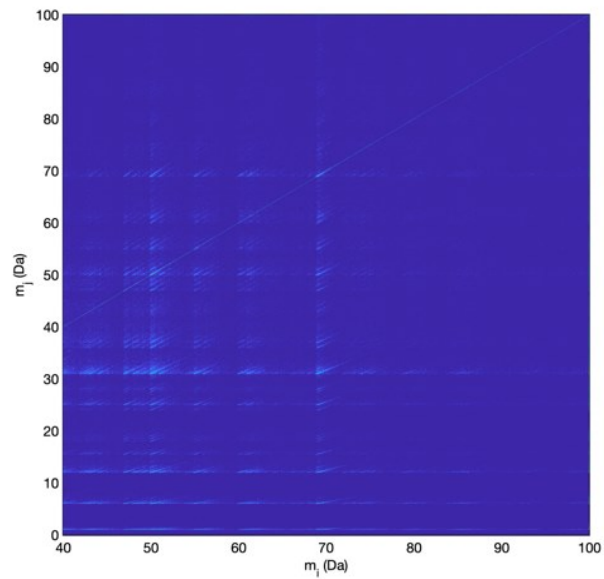

Figure S22 – Multiple hit correlation histogram of mass spectra from 40 Da to 100 Da collected from bulk polymer sharpened with room temperature Xe. Histogram shows no visible dissociation tracks.

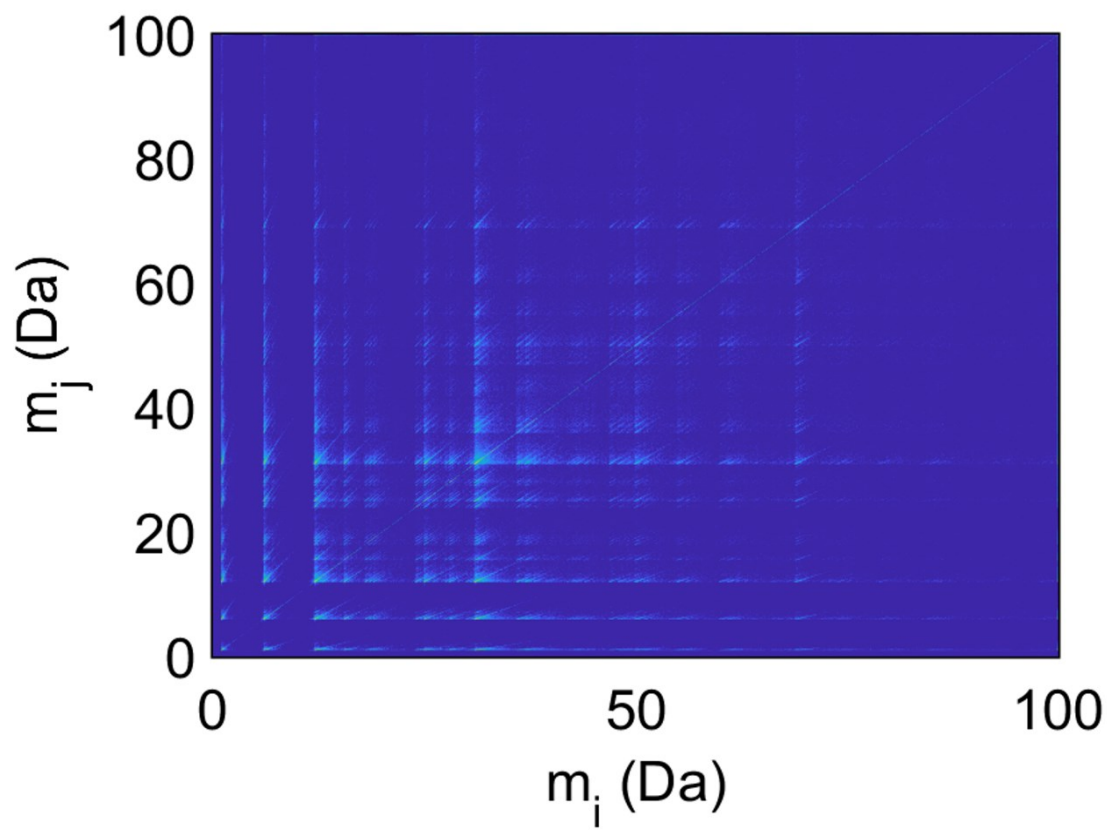

Figure S23 – Multiple hit correlation histogram of mass spectra up to 100 Da collected particle containing sample sharpened with room temperature Xe. Histogram shows no visible dissociation tracks.

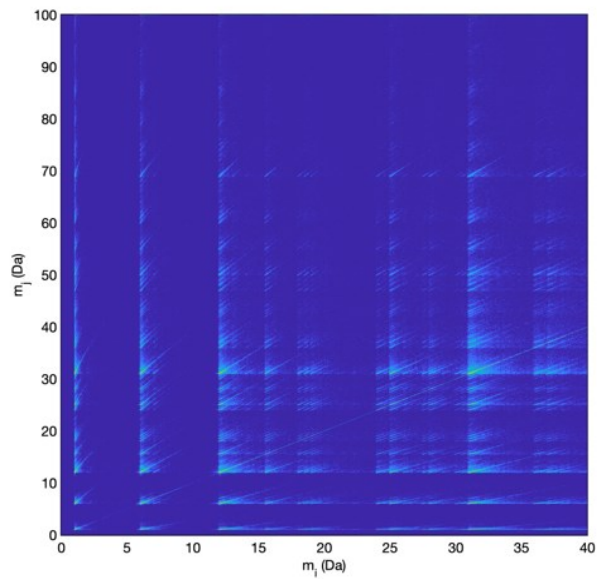

Figure S24 – Multiple hit correlation histogram of mass spectra from 0 to 40 Da collected from particle containing sample sharpened with room temperature Xe. Histogram shows no visible dissociation tracks.

c

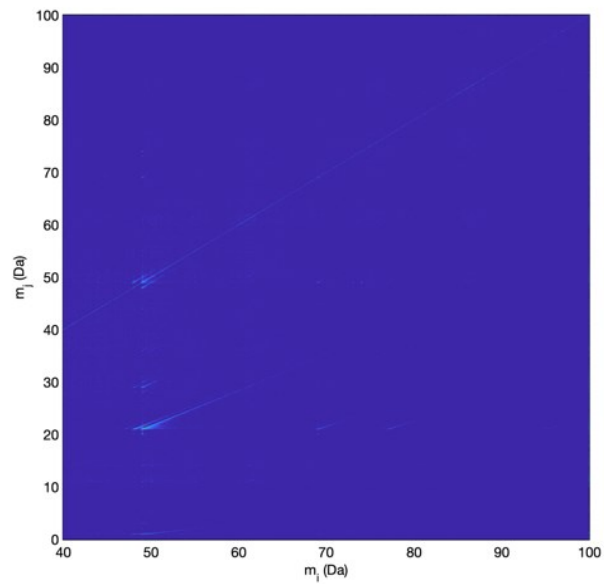

Figure S24 – Multiple hit correlation histogram of mass spectra from 0 to 40 Da collected from particle containing sample sharpened with room temperature Xe. Histogram shows no visible dissociation tracks.

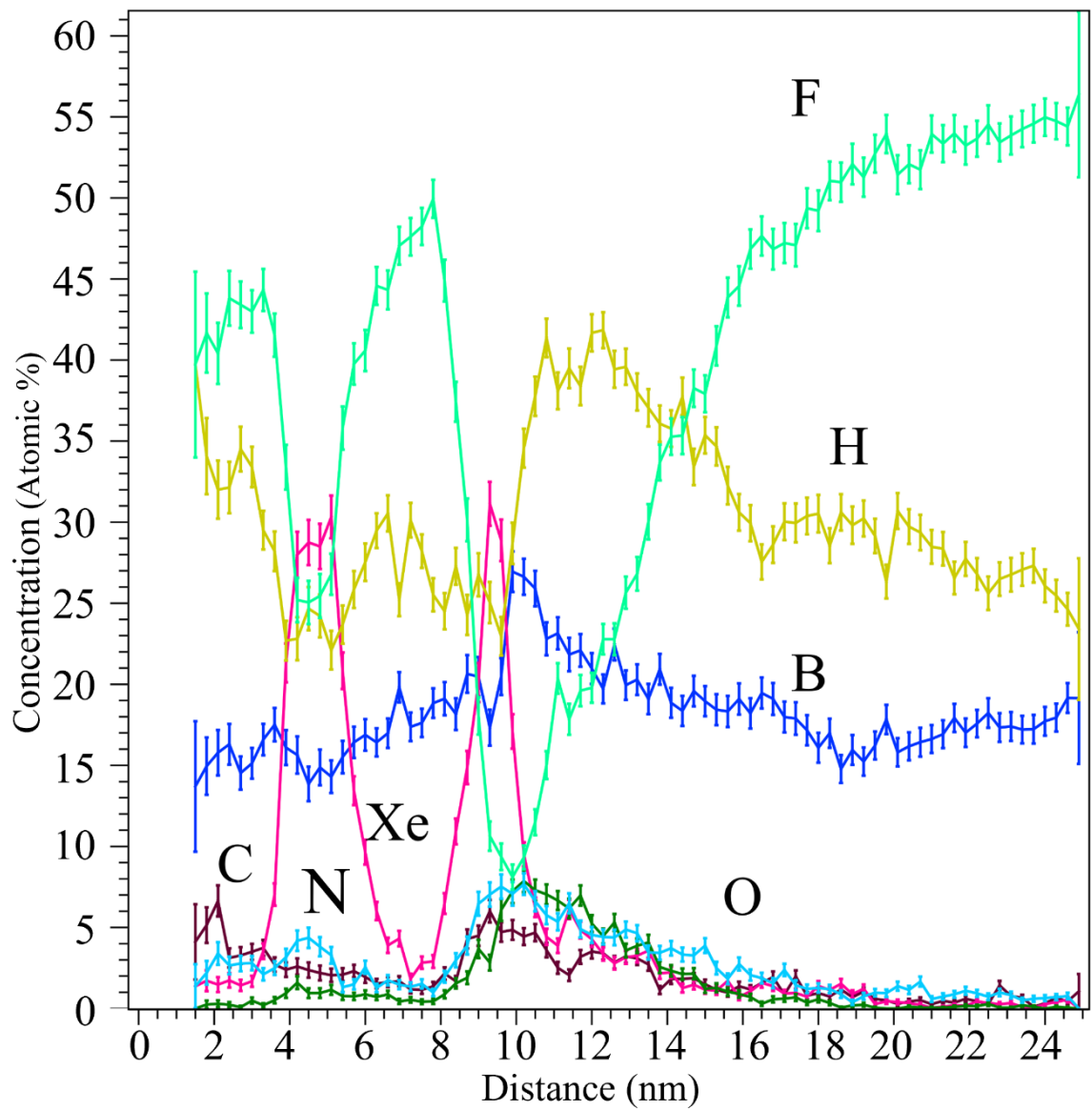

Figure S26 – 1D concentration profile of decomposed elements in analysis (Z) direction of the cylindrical region of interest shown in Figure 8. Profile generated from the full length of the sample. F species (green), H species (yellow), boron species including all potential boron peaks (dark blue), Xe species (pink), N species (green), C species (brown), O species (light blue)

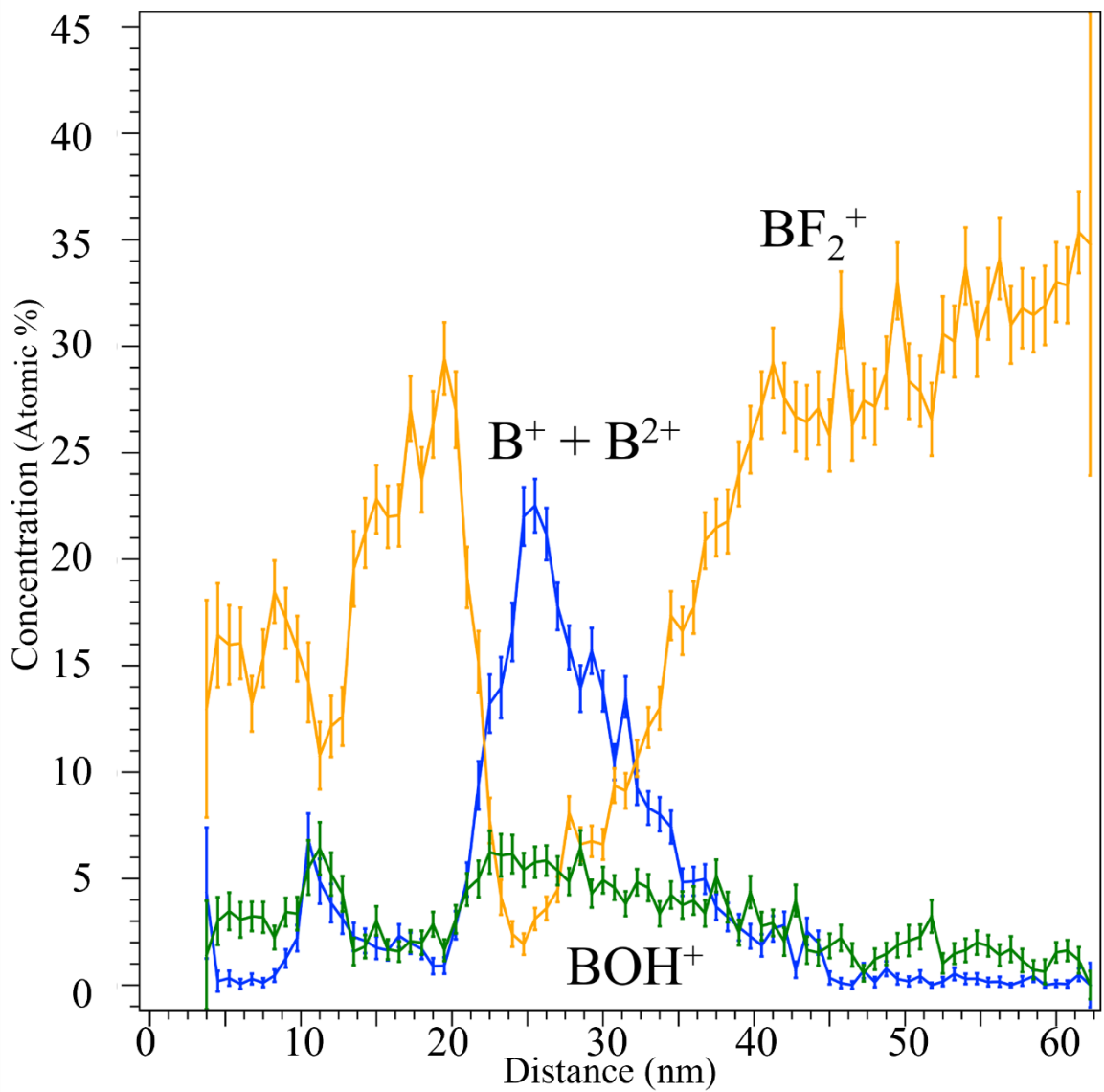

Figure S27 – 1D concentration profile of potential B species across cylindrical Region of Interest in Figure 8. The species identified as $B^+$ and $B^{2+}$ species and the $BOH^+$ species show a peak at the centre and so are assumed to be unambiguously B containing while the species labelled $BF_2^+$ containing a trough in the same region.

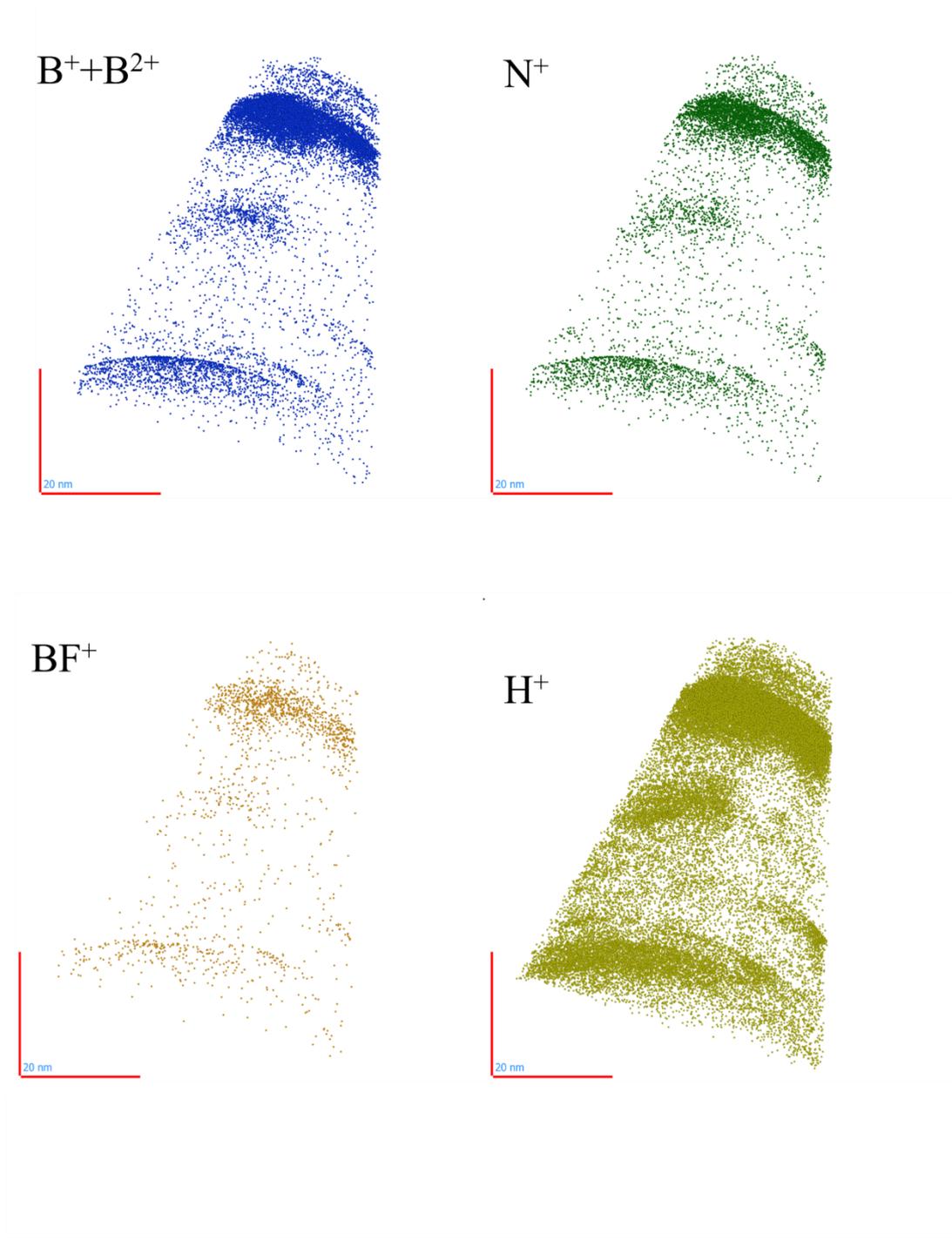

Figure S28 - An alternative angle for viewing the variation in some species of interest ($B^+$, $B^{2+}$, $N^+$, $BF^+$ and $H^+$) in the reconstruction of the room temperature Xe sharpened particle containing sample shown in Figure 8. Scale bar 20 nm.

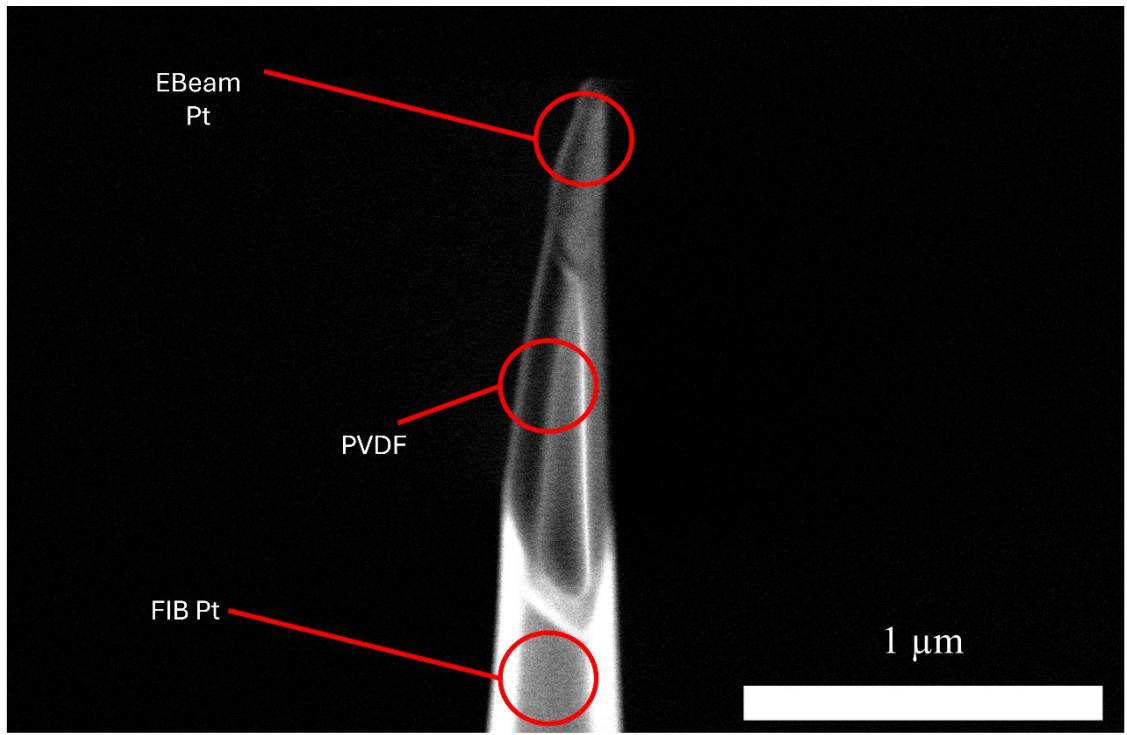

Figure S29 – SEM micrograph of a room temperature Xe sharpened bulk sample that shows visible bending under the electron beam, with beam damage likely occurring. The sample is not fully sharpened and retains a layer of protective Pt compound at the apex, indicating that beam damage may be occurring at stages of the sample preparation before the sample is fully sharpened. Red circles highlight regions of interest

| Sample preparation method | C (.at%) | F (.at%) | H (.at%) | O (.at%) |
|---|---|---|---|---|
| RT Xe | ~43 | ~30 | ~25 | ~2 |
| RT Ga | ~53 | ~22.5 | ~22.25 | ~2 |
| Cryo Ga | ~32 | ~43 | ~25 | ~1 |

Table 1 - The decomposed elemental plateau concentrations from room temperature Xe FIB sharpening (Figure 5B), room temperature Ga FIB sharpening (Supplementary Figure S18) and cryogenic Ga FIB sharpening (Supplementary Figure S19).